\pdfoutput=1
\documentclass[11pt]{article}

\usepackage[preprint]{acl}

\usepackage{times}
\usepackage{latexsym}

\usepackage[T1]{fontenc}
\usepackage[utf8]{inputenc}

\usepackage{microtype}

\usepackage{inconsolata}

\usepackage{graphicx}

\usepackage[most]{tcolorbox}

\usepackage[utf8]{inputenc} % allow utf-8 input
\usepackage[T1]{fontenc}    % use 8-bit T1 fonts
\usepackage{hyperref}       % hyperlinks
\usepackage{url}            % simple URL typesetting
\usepackage{booktabs}       % professional-quality tables
\usepackage{amsfonts}       % blackboard math symbols
\usepackage{microtype}      % microtypography
\usepackage{graphicx}
\usepackage{multicol}
\usepackage{multirow}
\usepackage{wrapfig}
\usepackage{pifont}
\usepackage{enumitem}
\usepackage{tcolorbox}
\usepackage{colortbl}
\usepackage{subcaption}

\lstdefinestyle{plain}{
    basicstyle=\fontsize{7}{9.5}\ttfamily,
    keywordstyle=\color{blue},
    commentstyle=\color{gray},
    stringstyle=\color{green},
    showstringspaces=false,
    breaklines=true,
    breakatwhitespace=false,
    breakindent=0pt,
    escapeinside={(*@}{@*)}
}

\renewcommand{\thefootnote}{\fnsymbol{footnote}}

\title{\ours: Evaluating LLM Agent’s Ability on \\ Reproducing \underline{L}anguage \underline{M}odeling \underline{R}esearch}

\renewcommand{\thefootnote}{\fnsymbol{footnote}}

\author{
{\bf Shuo Yan\footnotemark[1], Ruochen Li\footnotemark[1], Ziming Luo\footnotemark[1], Zimu Wang\footnotemark[1], Daoyang Li\footnotemark[1],} \\
{\bf Liqiang Jing, Kaiyu He, Peilin Wu, George Michalopoulos,} \\
{\bf Yue Zhang, Ziyang Zhang, Mian Zhang, Zhiyu Chen, Xinya Du} \\
University of Texas at Dallas \\
\texttt{\{shuo.yan, ruochen.li, ziming.luo, zimu.wang\}@utdallas.edu} \\
}

\newcommand{\ours}{\textsc{LMR-Bench}}

\begin{document}
\maketitle

\footnotetext[1]{Equal contribution.}

\setcounter{footnote}{0}  
\renewcommand{\thefootnote}{\arabic{footnote}}

\begin{abstract}
  Large language model (LLM) agents have demonstrated remarkable potential in advancing scientific discovery. However, their capability in the fundamental yet crucial task of reproducing code from research papers, especially in the NLP domain, remains underexplored. This task includes unique complex reasoning challenges in the intellectual synthesis of abstract concepts and the comprehension of code repositories with interdependent files.
  % \zhiyu{emphasize challenges on complex reasoning?}
  Motivated by this gap, we present \ours, a benchmark designed to systematically evaluate the capability of LLM agents on code reproduction from \textbf{\underline{L}}anguage \textbf{\underline{M}}odeling \textbf{\underline{R}}esearch. It consists of 28 code reproduction tasks derived from 23 research papers published in top-tier NLP venues over the past five years, spanning nine fundamental categories.
  Models are provided with a research paper, a code repository containing one or more masked functions,
  % \zhiyu{we also have data for more than one functions?},
  and instructions for implementing these functions.
  We conduct extensive experiments in standard prompting and LLM agent settings with state-of-the-art LLMs, evaluating the accuracy of unit tests and performing LLM-based evaluation of code correctness.
  Experimental results reveal that even the most advanced models still exhibit persistent limitations in scientific reasoning and code synthesis, highlighting critical gaps in LLM agents' ability to autonomously reproduce scientific research\footnote{Data, code, and leaderboard of \textsc{LMR-Bench} are available at \url{https://github.com/du-nlp-lab/LMR-Bench}.}. 
  % We will release our benchmark and code after publication.
\end{abstract}

% % Add footnote without symbol
% \begingroup
% \renewcommand\thefootnote{}
% \footnotetext{Data, code, and leaderboard at \url{https://github.com/du-nlp-lab/LMR-Bench}.}
% \addtocounter{footnote}{-1}
% \endgroup

\section{Introduction}
\label{sec:intro}

% \shuo{Introduce LLMs/Agents for scientific discovery}
The advent of large language model (LLM) agents has revolutionized beyond language generation, being recognized as a transformative force in advancing scientific discovery \citep{DBLP:conf/iclr/SiYH25, DBLP:journals/corr/abs-2311-08993,DBLP:conf/emnlp/MaGHXWPY0S24, DBLP:conf/acl/YangDLZPC24, DBLP:journals/corr/abs-2501-04306}. These agents have shown to be capable of executing the entire scientific discovery pipeline \citep{DBLP:journals/corr/abs-2503-22708, DBLP:journals/corr/abs-2504-08066, DBLP:journals/corr/abs-2408-14033}, from generating research ideas, designing experiments \citep{DBLP:journals/corr/abs-2410-13185,DBLP:journals/corr/abs-2412-14626,DBLP:conf/naacl/BaekJCH25} to implementing code \citep{DBLP:journals/corr/abs-2406-00515,DBLP:journals/tosem/JiangDWFSLJJ24,DBLP:conf/acl/ZhangLLSJ24}, drafting academic papers \citep{DBLP:conf/chi/0002LY22,DBLP:conf/nips/WangGYZZ0ZD0W0Z24,doi:10.1056/AIoa2400555}, and even producing complete papers that potentially pass peer review \citep{DBLP:journals/corr/abs-2504-08066}. They have also been integrated with tools such as Scholar Inbox \citep{DBLP:journals/corr/abs-2504-08385} to accelerate human's research, highlight the extraordinary capability of LLM agents to understand, synthesize, and generate complex knowledge in scientific discovery.

Despite the agents' remarkable advancement in research acceleration, there remains a notable gap regarding their ability in a foundational yet crucial aspect of scientific validation, i.e., code reproduction from academic papers in real-world environments.
This task poses unique challenges in complex reasoning, especially in the following two aspects: (1) \textbf{Logic understanding}, such as the intellectual synthesis of concise and abstract mathematical equations, algorithm outlines, and generalized flowchats; (2) \textbf{Code implementation}, particularly at the repository level that spreads across multiple interdependent files. Reproducing algorithms necessitates a thorough analysis of these complex dependencies, ensuring the consistency of both the internal codebase and the external environment, thereby amplifying the challenges associated with reproduction.
% \zhiyu{Maybe you should focus more on reasoning challenges on understanding and implementing research ideas and algorithms? I remember most of our data only need one file?}

\begin{figure*}[t!]
    \centering
    \includegraphics[width=\linewidth]{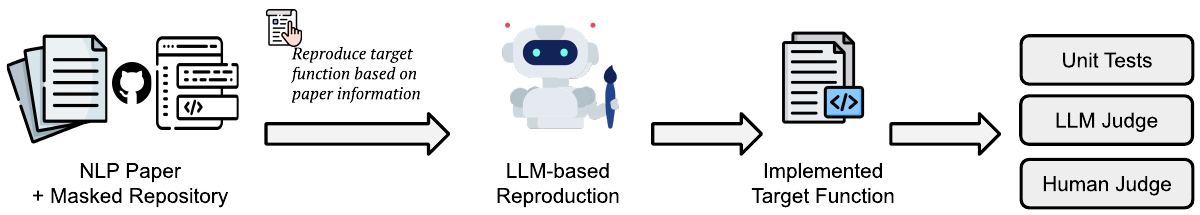}
    \vspace{-5mm}
    \caption{Overview of \ours. Given an NLP research paper and a corresponding codebase with masked functions, the LLM agent is tasked with reproducing the function, requiring its ability of scientific method understanding, abstract reasoning and cross-file understanding.} 
    \vspace{-4mm}
    \label{fig:example}
\end{figure*}

However, while reproducing code from research papers is a critical capability for LLMs, there is a lack of a dedicated benchmark that systematically evaluates the capability of LLMs to reproduce research papers in real-world scientific contexts. 
Existing efforts fall into several categories: ML engineering (e.g., MLAgentBench \citep{DBLP:conf/icml/HuangVLL24}, MLE-Bench \citep{DBLP:conf/iclr/ChanCJASMSLMPMW25}), data-driven scientific discovery (e.g., DSBench \citep{DBLP:conf/iclr/JingHWYYM0DY25}, ScienceAgentBench \citep{DBLP:conf/iclr/ChenCNZWYLLWLDX25}), and code debugging and issues resolving (e.g., SWE-bench \citep{DBLP:conf/iclr/JimenezYWYPPN24}, DebugBench \citep{DBLP:conf/acl/TianYQCLPWHL0024}). 
While these benchmarks are valuable, they typically evaluate isolated technical capabilities using task-specific inputs and metrics, rather than evaluate end-to-end paper reproduction.
A concurrent effort, PaperBench \citep{DBLP:journals/corr/abs-2504-01848}, evaluates LLM agents on the reproduction of 20 ICML 2024 papers. However, PaperBench requires reproducing the entire project from scratch—an unrealistic expectation for current LLM agents, making it difficult to offer valuable insights in guiding model improvements. More importantly, its evaluation protocol relies solely on LLM-as-a-judge, lacking curated unit tests or automated checks to ensure objective and reproducible assessments.

 % and may result in discrepancies when compared to human-curated codebases.

% However, this dataset requires reproduction from scratch, a task that significantly exceeds the current capabilities of existing agents. This limitation makes it challenging to extract meaningful insights for improving the models and may result in inconsistencies when compared to human-curated codebases.

% This gap motivates the need for benchmarks that better reflect the demands of real-world research workflows.
% These benchmarks differ substantially in their task objectives, data formats, and evaluation methods from research paper reproduction, thereby limiting the potential of LLMs to assist in the practical implementation and improve reproducibility standards within the field.
% \zhiyu{How do these datasets differ? List some core points corresponding to the challenges listed above.}

% \rl{To this end, we propose \textsc{NLPReproBench}, a benchmark that targets the core reproducibility challenges in NLP research, requiring models to translate abstract algorithm descriptions into correct, runnable code across real-world codebases.}\rl{NLPReproBench? AlgoReproBench}

Motivated by this gap, we present \ours, a benchmark designed to systematically evaluate the LLM agent's ability on reproducing \textbf{\underline{L}}anguage \textbf{\underline{M}}odeling \textbf{\underline{R}}esearch.
% \zhiyu{Why call agent-based? Both standalone and agent models can do it.} from NLP research papers.\shuo{will be revised by Shuo later. plan: We test both LLMs and Agent on repo-level code reproduction.}
It consists of 28 code reproduction tasks derived from 23 research papers published in top-tier NLP venues over the past five years, spanning nine fundamental categories, such as generative models and reinforcement learning, which are central to current LLM research.
In each task, LLM agents are provided with a research paper, a code repository containing one or more masked functions, and instructions for implementing these functions.
Successful completion requires the model to comprehend the algorithmic details accurately and generate functionally correct, syntactically consistent code (Figure \ref{fig:example}).
To ensure an objective evaluation of the code reproduction results, we design two distinct metrics: the accuracy of unit tests curated by human expert annotators and the distribution of LLM-as-a-judge classifications of generated implementations into three categories.

We conduct extensive experiments in standard prompting and LLM agent settings with OpenHands \citep{DBLP:conf/iclr/0001LSXTZPSLSTL25} on state-of-the-art LLMs, such as GPT-4o \cite{DBLP:journals/corr/abs-2410-21276}, GPT-4.1, and o4-mini. Experimental results reveal that even the most advanced models and LLM agents exhibit persistent limitations in scientific reasoning and code synthesis, such as unsuccessful paper parsing and failure in reasoning across steps and files, highlighting critical gaps in agent's ability to autonomously reproduce scientific research.
% \zhiyu{You can summarize a few core limitation types from your analysis.}

The key contributions of our work can be summarized as follows:

\begin{itemize}[itemsep=0pt,topsep=0pt,parsep=0pt]
    \item We present \ours, a benchmark designed to systematically evaluate the ability of LLM agents to reproduce scientific research projects. It consists of 28 code reproduction tasks across nine core categories in LLM research. 
    \item We introduce two complementary evaluation metrics: the accuracy of unit tests and the distribution of LLM-as-a-judge classifications of generated implementations, offering an objective evaluation of the LLM agent's capabilities. The unit tests are curated by human expert annotators and executed within separate Docker containers to faithfully reproduce the original environment.
    \item We conduct extensive experiments in standard prompting and LLM agent settings, highlighting critical gaps and producing valuable insights in current LLM agent’s ability in reproducing scientific research.
\end{itemize}

\section{Related Work}
\label{sec:related}

\paragraph{LLMs Agents for Scientific Discovery.}
% \shuo{LLMs/Agents?}

Recent research has increasingly focused on leveraging LLM agents to advance scientific discovery, spanning the entire research pipeline, from research idea generation and experimental design \citep{DBLP:conf/iclr/SiYH25, DBLP:journals/corr/abs-2410-13185,DBLP:journals/corr/abs-2412-14626,DBLP:conf/naacl/BaekJCH25} to implementing code \citep{DBLP:journals/corr/abs-2406-00515,DBLP:journals/tosem/JiangDWFSLJJ24,DBLP:conf/acl/ZhangLLSJ24} and drafting academic papers \citep{DBLP:conf/chi/0002LY22,DBLP:conf/nips/WangGYZZ0ZD0W0Z24,doi:10.1056/AIoa2400555}.
Some studies have introduced agent-based systems that can automate an end-to-end research flow. CodeScientist~\citep{DBLP:journals/corr/abs-2503-22708} and 
\textsc{MLR-Copilot}~\citep{DBLP:journals/corr/abs-2408-14033} utilize LLM agents to autonomously generate and implement research ideas. AI Scientist \citep{DBLP:journals/corr/abs-2408-06292} is an iterative framework that automates research idea generation, experimentation, paper writing, and peer review. AI Scientist-v2 \citep{DBLP:journals/corr/abs-2504-08066} extends the pipeline with a tree-search-based experiment management strategy, which produces manuscripts that pass peer review at leading machine learning workshops.
However, the ideas and experiments in these frameworks are typically synthesized by agent themselves, which limits their ability to capture the complexity of real-world scenarios. In contrast, our research focuses on the capability of LLM agents to faithfully reproduce peer-reviewed research papers, bridging the gap between agent-synthesized information and real-world publications.
% \shuo{Why do we need to mention the difference in related work? We can show these in the introduction part.}

\paragraph{LLM Agents for Code Generation.}

Code generation serves as a recognized benchmark for evaluating modes' problem-solving abilities and their practicality in software development. Models such as Codex~\citep{DBLP:journals/corr/abs-2107-03374} and Qwen-Coder~\citep{DBLP:journals/corr/abs-2409-12186}, accompanied by agent-based frameworks like MapCoder \cite{DBLP:conf/acl/IslamAP24}, AgentCoder~\cite{DBLP:journals/corr/abs-2312-13010},  and OpenHands~\cite{DBLP:conf/iclr/0001LSXTZPSLSTL25} have been proposed to improve the scalability of code intelligence. Meanwhile, systems like AIDE~\citep{DBLP:journals/corr/abs-2502-13138} and R\&D Agent~\citep{li2025rdagentquant} focus on automating machine learning engineering tasks through agentic planning.
To better benchmark these agents, some benchmarks dedicated for code generation have been proposed.
MBPP, MathQA-Python \citep{DBLP:journals/corr/abs-2108-07732}, FC2Code \citep{DBLP:conf/emnlp/LiuHZLZX22}, and LiveCodeBench \citep{DBLP:conf/iclr/JainHGLYZWSSS25} evaluate models' capability to generate code based on natural language instructions.
MLAgentBench \citep{DBLP:conf/icml/HuangVLL24} and MLE-bench \citep{DBLP:conf/iclr/ChanCJASMSLMPMW25} are based on Kaggle competitions to evaluate LLMs' machine-learning engineering capabilities.
RepoBench \citep{DBLP:conf/iclr/0003XM24} and ML-Bench \citep{DBLP:journals/corr/abs-2311-09835} focus on code generation at the repository level.
DSBench \citep{DBLP:conf/iclr/JingHWYYM0DY25} and SciAgentBench \citep{DBLP:conf/iclr/ChenCNZWYLLWLDX25} emphasize data-driven scientific discovery.
SWE-bench \citep{DBLP:conf/iclr/JimenezYWYPPN24} and DebugBench \citep{DBLP:conf/acl/TianYQCLPWHL0024} center on resolving bugs and issues within the codebase.
However, few benchmarks focus on \emph{research code implementation and execution}, which is a fundamental capability of LLM research agents \cite{DBLP:journals/corr/abs-2501-04306}.
PaperBench \cite{DBLP:journals/corr/abs-2504-01848} assesses LLM agents to replicate 20 research papers from ICML 2024. However, this dataset requires reproduction entirely from scratch, which is far beyond existing agents' ability and may lead to discrepancies compared to human-curated codebases.
In this paper, we build upon the concept of code generation but focus on specifically the exciting and fundamental skill of code reproduction from NLP research papers with human instructions.

\section{\ours}

\begin{figure}[t!]
    \centering
    \includegraphics[width=\linewidth]{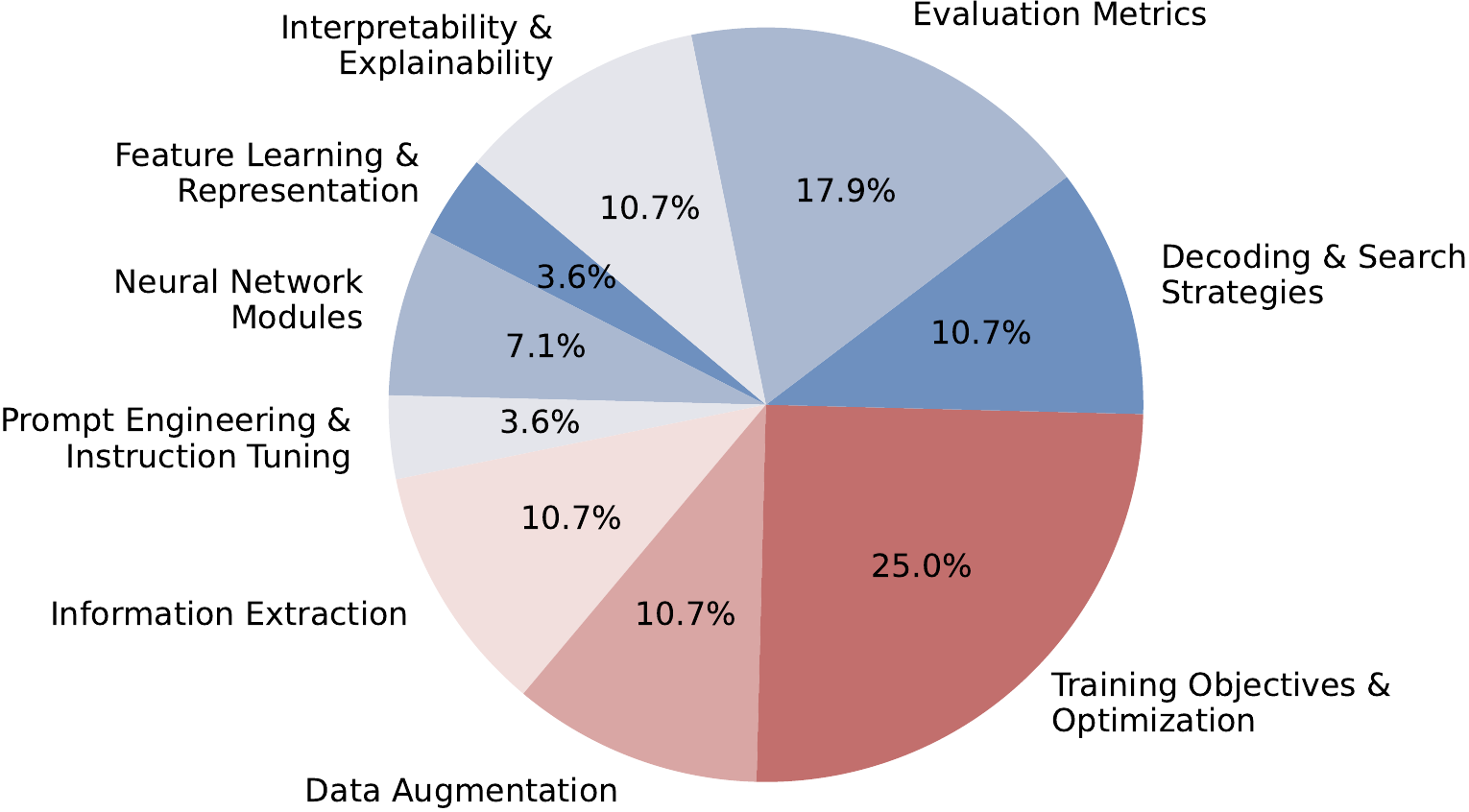}
    \caption{Question distribution in \textsc{\ours}.}
    \label{fig:distribution}
    \vspace{-5mm}
\end{figure}

\textsc{\ours} is a benchmark designed to evaluate the LLM agent's ability on reproducing language modeling research papers on related functions in the repository. 
It consists of 23 research papers and 28 distinct questions, each corresponding to a key task within the LLM/NLP research field. It covers nine essential task categories (see Table \ref{tab:categories}), with the distributions illustrated in Figure \ref{fig:distribution}.
Training Objectives \& Optimization and Evaluation Metrics are the most prevalent. This distribution aligns with the significance and practical challenges of real-world reproducibility,  as these areas often involve high levels of abstraction and require rigorous methodological precision.

\begin{figure*}[t!]
    \centering
    \includegraphics[width=\linewidth]{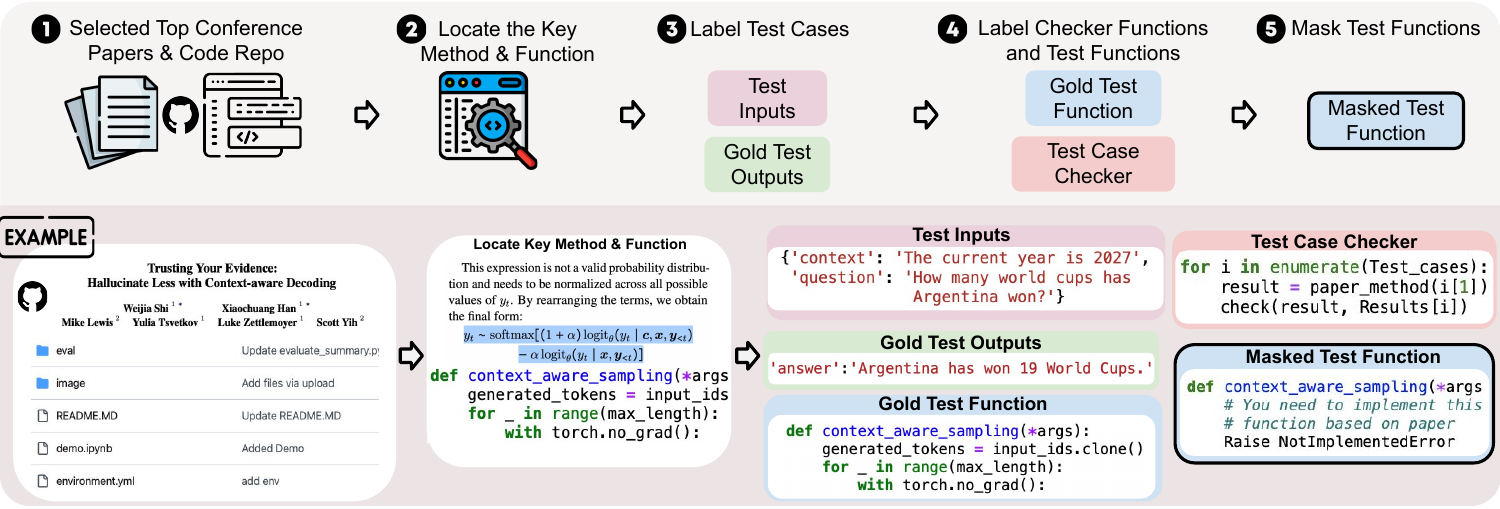}
    \caption{Dataset annotation pipeline of \textsc{\ours}.}
    \label{fig:annotation_pipeline}
    \vspace{-4mm}
\end{figure*}

\subsection{Task Formulation}

Given a research paper and its corresponding open-source code repository, the aim of the task is to reproduce the implementation of the missing function using the information provided in the paper, focusing on repository-level code generation. This simulates the real-world scenario where reproduce or verify key components of a research method based on its textual description.

As illustrated in Figure 1, the components provided as inputs to LLMs include (1) the original paper obtained from the proceedings; (2) the source files and codes within the repository, with masked functions for reproduction;
% \zhiyu{``Minor revision'' is confusing. Maybe just say masked method.},
and (3) the definition of the target function, including the detailed description on its definition, input, output, and any additional steps required for implementation.
% \shuo{Dataset should be refined after ddl.}
The reproduction process involves two different setup: standard prompting and LLM agent settings. The output function is evaluated via a combination of unit tests and LLM-as-a-judge method, offering a multi-faceted evaluation of correctness and fidelity.
% \shuo{We provide 2 automatic evaluation methods here and to verify the reliability of these evaluation methods, we also perform human evaluation on our settings.}

\subsection{Data Collection}
\label{sec:data_collection}

Figure \ref{fig:annotation_pipeline} illustrates the collection pipeline of \ours{}. In this section, we introduce each process in detail.

\paragraph{Paper Selection.}
% \zhiyu{Need details for how to gather the annotators and their background. Perhaps put into ethical statement (if permitted) and appendix? }
We form a group of 12 experienced researchers
% \zhiyu{Add (Our co-authors). You should also refer to ethical statement for more details, to avoid ？confusions and ethics review.}
in the LLM/NLP research community as our co-authors and annotators. Each annotator has been provided with a detailed, step-by-step annotation guideline, accompanied with examples.
% \zhiyu{Strictly speaking, you should provide the guidelines in appendix and ensure anonymous.}
We instruct annotators to collect research papers published within the past five years from top-tier NLP conferences, including ACL, EMNLP, NAACL, EACL, and COLING, and select appropriate papers for annotation. Each candidate paper must satisfy the following criteria:
(1) \textbf{Methodological Focus:} The paper should retain method-driven research rather than survey papers or benchmarks;
(2) \textbf{Reproducibility:} The proposed method should have an official, up-to-date repository with most issues resolved, ensuring that results are reproducible;
(3) \textbf{Clarity and Complexity:} The method should be well-documented, with clear instructions and sufficient implementation details provided in the paper, and should involve a level of complexity beyond basic examples (e.g., simple prompts like ``\textit{Let's think step by step}'').
Before the annotation process, we perform a manual review of the candidate papers selected by annotators to ensure their adherence to these criteria.

\begin{figure*}[t!]
    % \vspace{-30pt}
    \centering
    \includegraphics[width=\linewidth]{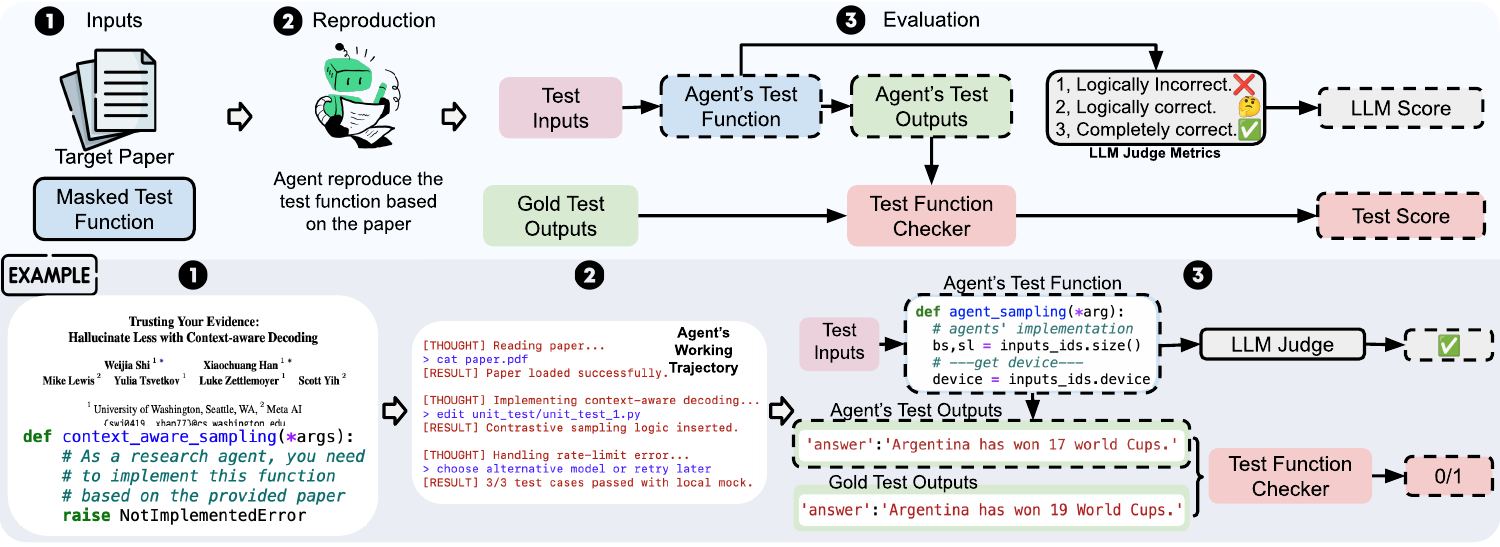}
    \caption{Dataset evaluation pipeline of \textsc{\ours}. The agent is presented with a target paper and a masked test function. After reproduction, the test function is evaluated in two stages. First, an LLM judge assesses the code for correctness and alignment with the paper’s logic (in this example, the judge deems the implementation correct). Second, the test function is executed on labeled inputs and its outputs are automatically compared against the golden outputs. In this figure, since there is only one test case, the final score is $0/1$.
    % (LLM judge results may vary with test cases.)
    }
    \label{fig:evaluation_pipeline}
    \vspace{-4mm}
\end{figure*}

\paragraph{Reproducibility Check.}

For each selected paper, annotators are required to reproduce its official codebase to guarantee its reproducibility. 
Environmental setup remains a persistent challenge, particularly due to the dependency conflicts across different repositories, especially those with proprietary components.
To mitigate this issue, we ask annotators to create a Dockerfile for each paper, following the repository's README. This ensures a consistent and functional execution environment.
More specifically, this process involves pulling an official PyTorch Docker image\footnote{\url{https://hub.docker.com/r/pytorch/pytorch}}, specifying the repository's dependencies, and incorporating any additional setup specific to the repository.
Annotators are required to resolve issues encountered in the environmental setup. Repositories with unresolved errors are excluded from the process.

\paragraph{Data Annotation.} During the data annotation process, annotators begin by selecting an algorithm presented in the paper and map it to the corresponding code block within the repository, ensuring the alignment between its theoretical description and code implementation. Next, for code blocks that are not organized into functions, they refactor the code into self-contained functions, ensuring them have appropriate inputs and outputs to encapsulate the core functionality.
Then, for each aligned function, annotators meticulously document essential implementation details in a structured format as comments, including its primary objective, inputs and outputs, intra- and inter-file dependencies, and additional steps required for implementation (e.g., usage of external APIs).
Following this documentation process, annotators craft detailed task instructions describing the algorithm's intended behavior. Subsequently, they generate a masked version of the function and save a golden file that will be utilized for evaluation.

% \shuo{Revise this part into one paragraph}
% \paragraph{Algorithm-Method Alignment.}

% In the first step, annotators begin by selecting an algorithm presented in the paper and map it to the corresponding code block within the repository, ensuring the alignment between its theoretical description and code implementation. Subsequently, they refactor the code into a self-contained method that encapsulates the core functionality. In cases where the code is already implemented as a method, annotators revise variable names and restructure code blocks to mitigate any potential risk of direct data leakage.\shuo{We did not do this now. Data leakage may be found in the experiments. Need to analyze the data now.}

% \paragraph{Detailed Annotation}

% For each aligned method, annotators meticulously document essential implementation details in a structured format as method comments, including its primary objective, inputs and outputs, intra- and inter-file dependencies, and additional steps required for implementation (e.g., usage of external APIs).
% Following this documentation process, annotators craft detailed task instructions describing the algorithm's intended behavior. Subsequently, they generate a masked version of the method and save a golden file that will be utilized for evaluation.

\paragraph{Unit Test Evaluation Preparation.}
Finally, annotators construct a unit test suite comprising around 3 test cases derived from the original datasets, where both the inputs and corresponding expected outputs are faithfully recorded during the reproduction phase.
\cite{DBLP:conf/coling/ZhaoLT0YL025,DBLP:journals/corr/abs-2504-01848}.
For scalable and consistent evaluation, annotators create task-specific evaluation scripts based on these pre-defined metrics that measure output accuracy. LLMs are leveraged to support the creation of unit tests to enhance efficiency.
To account for the inherent variability that may present in NLP implementations, we design checker functions specific to each task.
For instance, we evaluate the value differences between predictions and ground truths for optimization tasks, while setting a threshold for the BERTScore \cite{DBLP:conf/iclr/ZhangKWWA20} in prompt engineering tasks.
Following the annotation process, all annotations undergo rigorous human review and refinement to guarantee correctness and reproducibility.
% \zhiyu{If you have time, you should include a few full data examples in appendix, e.g., the target functions and unit tests. }

\begin{table*}[t!]
    \centering
    \small
    
    \begin{tabular}{l|cccccl}
        \toprule
        \textbf{Benchmark} & \textbf{Pub.} & \textbf{Repo.} & \textbf{Unit} & \textbf{Docker} & \textbf{Source} & \textbf{Task} \\
        \midrule
        MLE-bench \cite{DBLP:conf/iclr/ChanCJASMSLMPMW25} & \textcolor{BrickRed}{\ding{55}} & \textcolor{BrickRed}{\ding{55}} & \textcolor{BrickRed}{\ding{55}} & \textcolor{BrickRed}{\ding{55}} & Kaggle & Machine Learning Engineering \\
        MLAgentBench \cite{DBLP:conf/icml/HuangVLL24} & \textcolor{BrickRed}{\ding{55}} & \textcolor{ForestGreen}{\ding{51}} & \textcolor{ForestGreen}{\ding{51}} & \textcolor{BrickRed}{\ding{55}} & Kaggle & Machine Learning Engineering \\
        \midrule
        RepoBench \cite{DBLP:conf/iclr/0003XM24} & \textcolor{BrickRed}{\ding{55}} & \textcolor{ForestGreen}{\ding{51}} & \textcolor{BrickRed}{\ding{55}} & \textcolor{BrickRed}{\ding{55}} & GitHub & Code Auto-Completion \\
        ML-Bench \cite{DBLP:journals/corr/abs-2311-09835} & \textcolor{BrickRed}{\ding{55}} & \textcolor{ForestGreen}{\ding{51}} & \textcolor{ForestGreen}{\ding{51}} & \textcolor{ForestGreen}{\ding{51}} & GitHub & Code Auto-Completion \\
        \midrule
        SWE-bench \cite{DBLP:conf/iclr/JimenezYWYPPN24} & \textcolor{BrickRed}{\ding{55}} & \textcolor{ForestGreen}{\ding{51}} & \textcolor{BrickRed}{\ding{55}} & \textcolor{BrickRed}{\ding{55}} & GitHub & Resolve GitHub Issues \\
        DeBugBench \cite{DBLP:conf/acl/TianYQCLPWHL0024} & \textcolor{BrickRed}{\ding{55}} & \textcolor{BrickRed}{\ding{55}} & \textcolor{ForestGreen}{\ding{51}} & \textcolor{BrickRed}{\ding{55}} & LeetCode & Resolve Code Bugs  \\
        \midrule
        DSBench \cite{DBLP:conf/iclr/JingHWYYM0DY25} & \textcolor{BrickRed}{\ding{55}} & \textcolor{BrickRed}{\ding{55}} & \textcolor{ForestGreen}{\ding{51}} & \textcolor{BrickRed}{\ding{55}} & Kaggle & Data-Driven Discovery \\
        ScienceAgentBench \cite{DBLP:conf/iclr/ChenCNZWYLLWLDX25} & \textcolor{ForestGreen}{\ding{51}} & \textcolor{BrickRed}{\ding{55}} & \textcolor{ForestGreen}{\ding{51}} & \textcolor{BrickRed}{\ding{55}} & Research & Data-Driven Discovery \\
        \midrule
        PaperBench \cite{DBLP:journals/corr/abs-2504-01848} & \textcolor{ForestGreen}{\ding{51}} & \textcolor{ForestGreen}{\ding{51}} & \textcolor{BrickRed}{\ding{55}} & \textcolor{ForestGreen}{\ding{51}} & Research & Reproduce ICML Papers \\
        \ours & \textcolor{ForestGreen}{\ding{51}} & \textcolor{ForestGreen}{\ding{51}} & \textcolor{ForestGreen}{\ding{51}} & \textcolor{ForestGreen}{\ding{51}} & Research & Reproduce LLM/NLP Papers \\
        \bottomrule
    \end{tabular}
    \caption{Comparison between \ours{} and existing benchmarks.}
    \vspace{-3.5mm}
    \label{tab:comparison}
\end{table*}

\subsection{Evaluation}

The overall evaluation pipeline is illustrated in Figure~\ref{fig:evaluation_pipeline}.  Since evaluating code quality involves complementary aspects of correctness and robustness, we employ two automated methods: (1) unit‐test evaluation and (2) LLM‐as‐a‐judge evaluation.

\paragraph{Unit Test Evaluation.}
% Similar to the evaluation method used in LeetCode, we test the correctness of the implementation within a unit test framework. In reproducibility check and unit test evaluation preparation part of section \ref{sec:data_collection}, we already showed the details of how we prepared for the data of unit test. \zhiyu{Simplify this sentence, e.g., the details of creating unit tests are shown in section 3.2.} In the evaluation phase, we just create a container based on the dockerfile we built before and then perform the unit test in the container. Then we calculate the accuracy based on the pass rate of the unit test in the question level.
We first measure functional correctness using an automated unit‐test framework, following a protocol similar to LeetCode’s \cite{DBLP:journals/corr/abs-2311-09835, DBLP:conf/coling/ZhaoLT0YL025}.  As described in Section~\ref{sec:data_collection}, we generate a suite of test cases and create a container as the testing environment.  During evaluation, each candidate implementation is executed inside its own container, and we run the predefined test suite. We then compute the unit‐test accuracy as the fraction of problems for which all associated test cases pass.

\paragraph{LLM-as-a-Judge Evaluation.}
% The above unit test evaluation may not reflect the helpfulness of the generated code. \zhiyu{helpfulness is vague. Breify explain why unit tests cannot give us 100\% comprehensive evaluations.} So we also consider LLM-as-a-judge evaluation since passing the unit test requires high accuracy of the code generated \zhiyu{Rewrite this sentence. ``high accuracy of the code generated'' is something we definitely want! You need to explain why unit tests are not comprehensive, what is missing, e.g., unit tests only assess final correctness and cannot give fine-grained assessments in the middle ground which provides more insights for model performances}. We directly compare the file generated and the reference solution file with a judge LLM. We let the judge LLM analyze the differences between the two files and then classify the situation based on two criterion: method correctness and code implementation correctness. Finally each implementation can be classified into one of the three categories: incorrect method implementation, correct method implementation but subtle errors on code implementation and totally correct implementation. Then we calculate the percentage of each category as the final index.

Unit tests ensure basic functionality but cannot capture code readability, style conformance, or subtle semantic discrepancies that do not trigger test failures \cite{DBLP:conf/emnlp/TongZ24,DBLP:journals/corr/abs-2504-01848}.  To obtain a more granular assessment, we introduce LLM-as-a-judge as the second evaluation method. We present both the model‐generated function and the reference solution to the LLM judge and evaluate the generated code from two distinct but complementary perspectives: algorithmic logic correctness and implementation correctness. The algorithmic logic correctness evaluation verifies that the algorithm’s underlying mathematical design and theoretical logic are conceptually sound and consistent with the intended methodology, ensuring that for each valid input the algorithm would produce an output meeting its formal specification. In parallel, the implementation correctness evaluation scrutinizes the code to ensure it faithfully realizes the intended algorithmic logic. This involves checking that the code’s procedures and data handling strictly follow the algorithm’s design and that it robustly handles edge cases (e.g., empty inputs or unexpected input formats), uses appropriate data types, and behaves reliably at runtime.

Based on the combined outcomes of these two evaluations, we classify the generated code’s correctness into three categories: 
\begin{enumerate}[label=(\roman*),itemsep=0pt,topsep=0pt,parsep=0pt,leftmargin=*]
  \item \emph{Logically Incorrect}: the code’s foundational logic is flawed, rendering it incapable of producing correct results even with a perfect implementation;
  \item \emph{Logically Correct but Incorrectly Implemented}: the design of the logic is sound in principle, but the implementation fails to realize that design accurately (for instance, due to bugs or improper handling of edge cases);
  \item \emph{Completely Correct}: the logic is conceptually sound and its implementation in code is faithful and error-free, satisfying all specified requirements.
\end{enumerate}

We report the percentage of implementations falling into each category to provide a more fine-grained performance analysis of the agent's implementation accuracy.

\begin{table*}[t!]
    \centering
    \small
    \resizebox{\textwidth}{!}{
    \begin{tabular}{l|c|ccc|ccc}
        \toprule
        \multirow{2}{*}{\textbf{Model}} & \textbf{Unit Test} & \multicolumn{3}{c|}{\textbf{LLM-as-a-judge Evaluation}} & \multicolumn{3}{c}{\textbf{Human Evaluation}} \\
        \cmidrule{3-8}
         & \textbf{Accuracy} & Completely Correct & Logically Correct & Logically Incorrect & Completely Correct & Logically Correct & Logically Incorrect \\
        \midrule
        GPT-4o & 39.3\% & 17.9\% & 10.7\% & 71.4\% & 25\% & 14.3\% & 60.7\% \\
        GPT-4.1 & 42.9\% & 7.1\% & 28.6\% & 64.3\% & 32.1\% & 28.6\% & 39.3\% \\
        o4-mini & 42.9\% & 25\% & 21.4\% & 53.6\% & 35.7\% & 32.1\% & 32.1\% \\
        \midrule
        OpenHands (GPT-4o) & 25\% & 7.1\% & 7.1\% & 85.7\% & 21.4\% & 17.9\% & 60.7\% \\
        OpenHands (GPT-4.1) & 32.1\% & 17.9\% & 14.3\% & 67.9\% & 32.1\% & 25\% & 42.9\% \\
        OpenHands (o4-mini) & 35.7\% & 35.7\% & 14.3\% & 50\% & 39.3\% & 21.4\% & 39.3\% \\
        \bottomrule
    \end{tabular}
    }
    \caption{Evaluation results for standard prompting and LLM agent settings.}
    \vspace{-1.5mm}
    \label{tab:model-performance}
\end{table*}

\subsection{Comparison with Existing Benchmarks}

Table \ref{tab:comparison} presents a systematic comparison between \ours{} and nine existing benchmarks across four essential dimensions: derived from published research papers (\textbf{Pub.}), repository-level operation (\textbf{Repo.}), standard unit tests (\textbf{Unit}), and task-specific Docker environments (\textbf{Docker}).
From the table, it is evident that \ours{} is the only benchmark combining all four features,  distinguishing it as a uniquely robust benchmark in contemporary LLM/NLP research.

% \subsection{Data contamination}
% To mitigate the data contamination issues, we required the annotators to 

\section{Experiments}
\label{sec:experiments}

\subsection{Experimental Setup}

Our experiments are conducted under two settings: \textit{standard prompting} and \textit{LLM agent} settings, where backbone LLMs used include GPT-4o \cite{DBLP:journals/corr/abs-2410-21276}, GPT-4.1\footnote{\url{https://openai.com/index/gpt-4-1/}}, and o4-mini\footnote{\url{https://openai.com/index/introducing-o3-and-o4-mini/}}, serving as representative models for general-purpose, complex, and reasoning-oriented tasks, respectively.
% These models are evaluated on their ability to reproduce research papers under both standard prompting and LLM agent settings, forming the backbone of the OpenHands framework. \zhiyu{This paragraph is not well-organized. Need significant revision. You should first say we evaluate both standalone LLM setting and agent setting, then say we use these LLMs for both setting, then say for agent setting we use openhands.}

\paragraph{Standard Prompting.}
In the \textit{standard prompting} setting, since LLMs cannot directly process an entire repository as input, we design a straightforward pipeline for data pre-processing, extracting relevant information by parsing the paper and presenting the associated code, which is then formatted into a prompt and passed to the LLM (see Appendix \ref{appendix:prompt-nov}). Specifically, the prompt includes the JSON format of the paper parsed by PyMuPDF\footnote{\url{https://github.com/pymupdf/PyMuPDF}}, the code of the goal file, the instruction and the related code snippet in other files in the repository.

\paragraph{LLM Agent Setting.} In contrast, the \textit{LLM agent} setting aims for an end-to-end problem-solving approach. Here, the agent is provided with a folder consisting of the paper's PDF file and the code repository with masked functions as input, tasked with addressing the problem using any available method. OpenHands \cite{DBLP:conf/iclr/0001LSXTZPSLSTL25}, a well-known coding agent, meets these requirements. Under this setup, the objective is to allow the agent to handle as much of the task as possible, with minimal intervention, i.e., only providing an instruction specifying which function needs to be implemented.

\subsection{Results and Analysis}

% Put results for No Agent setting and OpenHands setting in different tables.

% \subsection{Action Analysis for OpenHands Agents}
\begin{figure*}[t!]
    \centering
    \includegraphics[width=\linewidth]{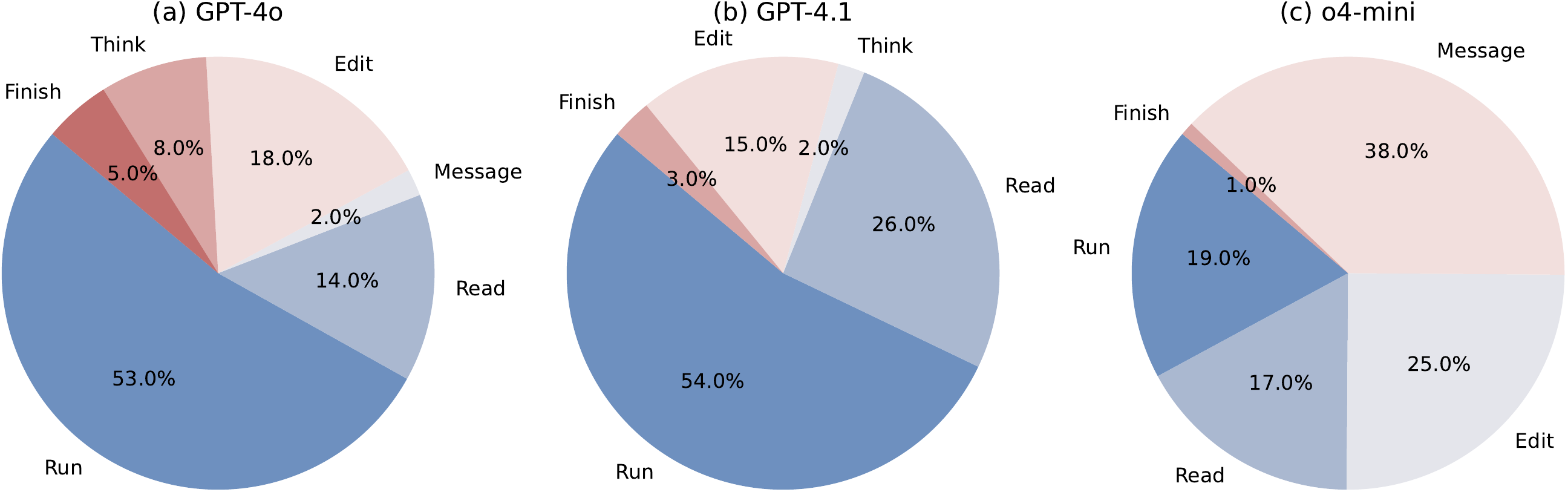}
    % \begin{subfigure}[b]{0.32\linewidth}
    %     \includegraphics[width=\textwidth]{latex/figures/GPT-4o.pdf}
    %     \caption{GPT-4o}
    %     \label{fig:gpt-4o}
    % \end{subfigure}
    % \hfill
    % \begin{subfigure}[b]{0.31\linewidth}
    %     \includegraphics[width=\textwidth]{latex/figures/GPT-4.1.pdf}
    %     \caption{GPT-4.1}
    %     \label{fig:gpt-4.1}
    % \end{subfigure}
    % \hfill
    % \begin{subfigure}[b]{0.35\linewidth}
    %     \includegraphics[width=\textwidth]{latex/figures/o4-mini.pdf}
    %     \caption{o4-mini}
    %     \label{fig:o4-mini}
    % \end{subfigure}
    \caption{Action distribution in OpenHands agents with different backbone models.}
    \vspace{-4mm}
    \label{fig:action_counts}
\end{figure*}

Table \ref{tab:model-performance} presents the overall performance comparison between standard prompting and LLM agents. GPT-4.1 and o4-mini achieve the highest accuracy on unit tests. However, when it comes to LLM-as-a-judge, o4-mini significantly outperforms GPT-4.1 in the number of samples deemed correct. On the other hand, GPT-4o exhibits the weakest performance, underscoring its limits in code reproduction. Although the absolute unit-test accuracy remains relatively low across all models, the category \textit{logically correct} also reflects the algorithms match the specification but contain implementation mistakes or omissions. This emphasizes the necessities of future models with enhanced abstract reasoning and better cross-file integration capabilities. 
% \zhiyu{What is the definition for ``almost correct''? I remember it means method is correct but with minor coding errors? Make sure the conclusions are consistent with your definitions.}

Compared to standard prompting, LLM agents show a slight decrease in accuracy across all models, with reductions of 14.3\%, 10.8\%, and 7.2\% for GPT-4o, GPT-4.1, and o4-mini, respectively. However, these agents tend to produce a higher number of functions identified as correct. This observation underscores the enhanced ability of LLM agents to understand paper details and generate accurate functions, while also revealing their limitations in repository-level paper reproduction that passes unit tests. Challenges such as repository-level code understanding and dependency handling emerge as key areas for improvement, offering valuable directions for future research.

% \paragraph{Performance Across Categories.}
% \shuo{We dont need that because we dont have much data for each category}

% \begin{table*}[t!]
%     \centering
%     \small
%     \caption{Performance across categories.}
%     \begin{tabular}{l|ccccccccc}
%         \toprule
%         \textbf{Model} & \textbf{FR} & \textbf{NN} & \textbf{PI} & \textbf{IE} & \textbf{DA} & \textbf{TO} & \textbf{DS} & \textbf{EM} & \textbf{INE} \\
%         \midrule
%         GPT-4.1 & 0\% & 50\% & 50\% & 33.3\% & 50\% &  \\
%         Claude-Sonnet-3.7 \\
%         Gemini-2.5 \\
%         DeepSeek-V3 \\
%         o4-mini \\
%         DeepSeek-R1 \\
%         OpenHands \\
%         \bottomrule
%     \end{tabular}
%     \label{tab:my_label}
% \end{table*}
% \shuo{We dont need this table.}

\section{Further Analysis}

% To understand how the agent's actions are related to the performance, we further analyze the actions of OpenHands agent. We find that different behaviour patterns exist between foundation models and reasoning models as shown in figure \ref{fig:action_counts}. From this figure, we find that the behaviour patterns between GPT-4o and GPT-4.1 are similar but GPT-4.1 prefers doing more actions instead of analyzing with chat. We also find the action patterns for reasoning models such as o4-mini are very different from foundation models. Reasoning models tend to do a lot of analysis and may 

% Although GPT-4.1 and o4-mini show different behaviour patterns, they achieve similar performance if we consider the results of human evaluation. We also find that the behaviours patterns between passed questions and failed questions share similar patterns in figure \ref{fig:} regardless of the type of the model.

\subsection{Action Analysis for OpenHands Agents}

To examine whether the distribution of actions of the agent is related to implementation success, we analyze the logs of OpenHands (Figure \ref{fig:action_counts}). Foundation models (GPT-4o, GPT-4.1) and reasoning models (o4-mini) exhibit markedly different interaction profiles. Although GPT-4o and GPT-4.1 share a similar overall action distribution, GPT-4.1 performs more concrete operations (e.g., code execution, file queries) and fewer conversational “think” steps. By contrast, the reasoning-oriented o4-mini devotes a larger fraction of its workflow to in-depth analysis before invoking execution steps.

\begin{figure}[t!]
    \centering
    \includegraphics[width=1\linewidth]{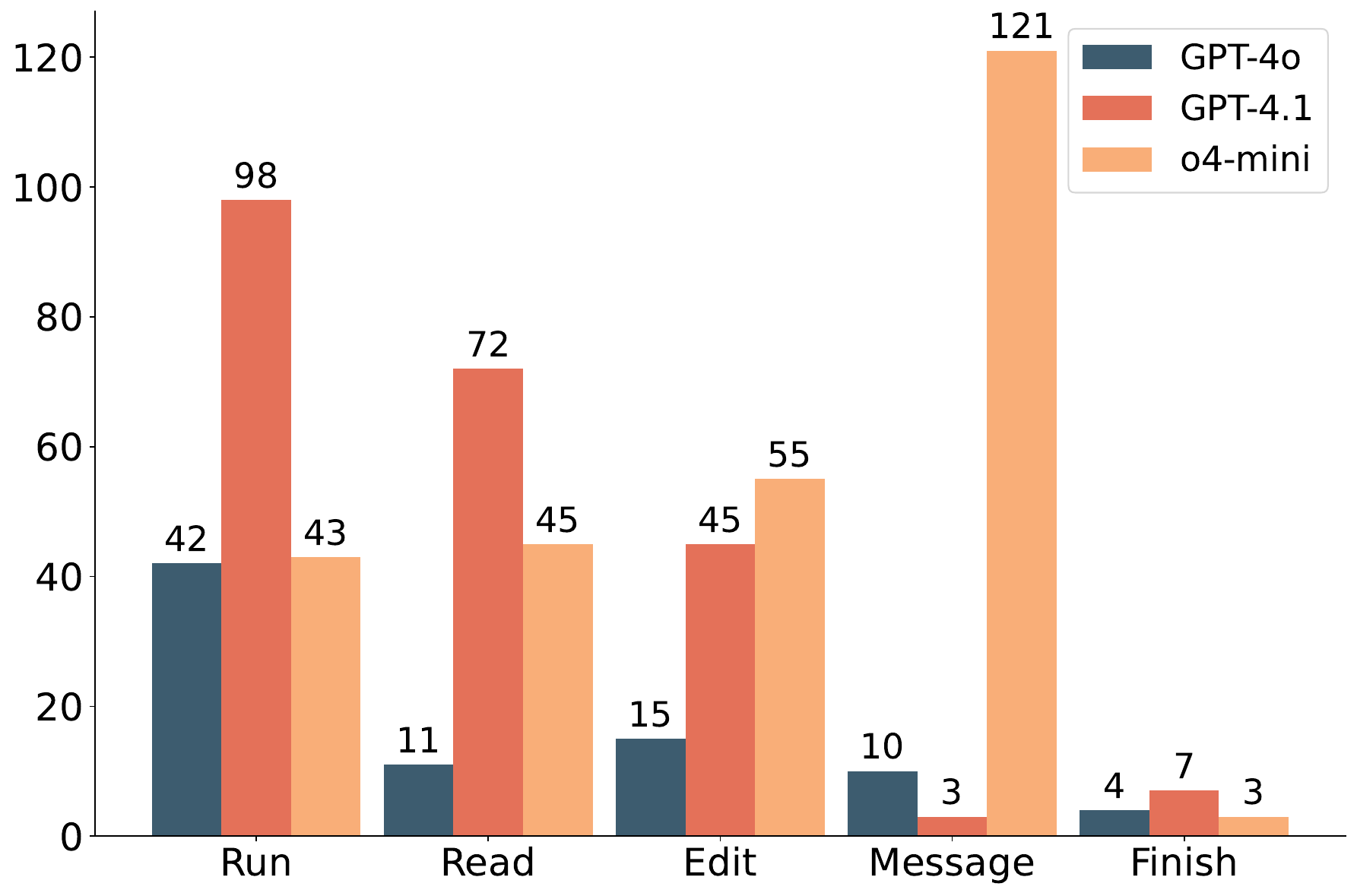}
    \includegraphics[width=1\linewidth]{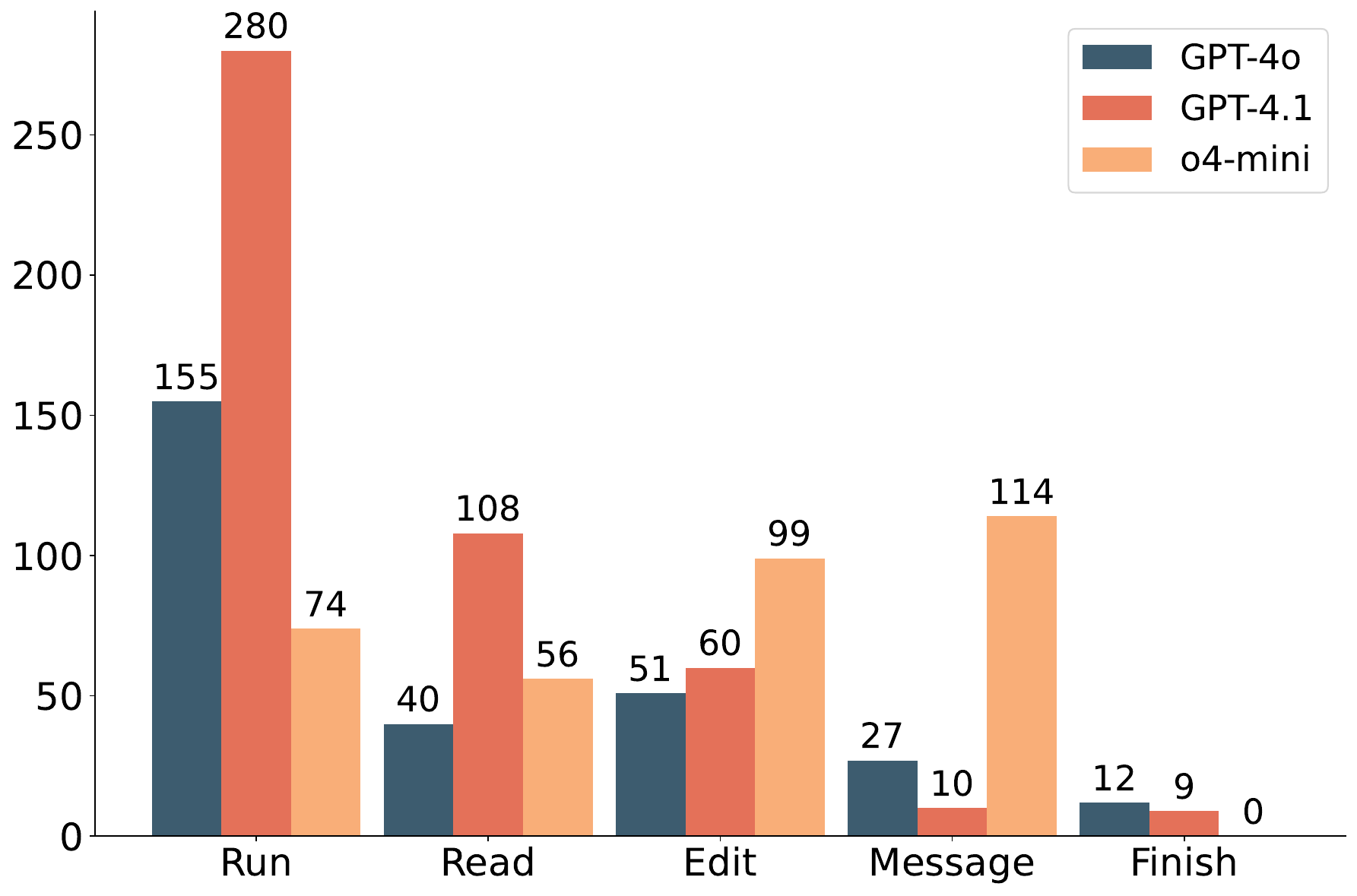}
    \vspace{-7mm}
    \caption{Action counts comparison for pass papers (\textit{upper figure}) and fail papers (\textit{bottom figure}).}
    \label{fig:action_counts_pass_fail}
    \vspace{-4.5mm}
\end{figure}

Despite these behavioral differences, GPT-4.1 and o4-mini achieve comparable success rates under human evaluation. Moreover, when we split logs by outcome (passed vs. failed), the relative balance between analysis and execution remains consistent across all model types (see Figure \ref{fig:action_counts_pass_fail}), suggesting that it is the ratio of “think” to “run” actions, rather than their absolute counts, that best predicts successful code implementation. The data analysis shows that the action distribution are not related to success and fail of the unit test since o4-mini, GPT-4.1 have different behavior pattern but show similar performance.

% \begin{figure}
%     \centering
%     \includegraphics[width=1\linewidth]{latex//figures/repo_analysis.png}
%     \caption{Logistic regression results}
%     \label{fig:repo_analysis}
% \end{figure}

To quantify the impact of repository structure on implementation success, we fit a logistic regression model (Table \ref{tab:logit-results} in Appendix \ref{app:logistic_analysis}) with binary success as the dependent variable, as predictors we included:

\begin{itemize}[itemsep=0pt,topsep=0pt,parsep=0pt,leftmargin=*]
    \item \textit{Average Directory Depth}: Sum the depth of each file or folder (where depth is the number of edges from the repository root to that node) and divide by the total number of nodes;
    \item \textit{Average Branch Factor}: Sum the number of immediate subdirectories for each directory node and divide by the number of directories with at least one child;
    \item \textit{Goal Function Length}: The number of source lines of code in the goal function file;
    \item \textit{Dummy Indicators for Model Type}: The current model is GPT-4.1 and two dummy variables to represent GPT-4o and o4-mini.
\end{itemize}

% * **Average directory depth** (\texttt{avg\_dir\_depth}),
% * **Average branch factor** (\texttt{avg\_branch\_factor}),
% * **Directory imbalance**, defined as the ratio of maximum to average depth (\texttt{directory\_imbalance}),
% * **Goal function length** in lines of code (\texttt{goal\_function\_length}),
% * Dummy indicators for model type (\texttt{model\_name\_gpt4o}, \texttt{model\_name\_o4mini}).

The regression reveals that deeper directory hierarchies are strongly associated with higher odds of success ($\beta = 0.8049$, $z = 3.232$, $p = 0.001$), whereas more highly branched structures significantly reduce success likelihood ($\beta = \text{–}0.6750$, $z = \text{–}2.995$, $p = 0.003$).  Neither directory imbalance nor goal‐function length reached statistical significance (both $p > 0.10$), nor did the model‐type indicators ($p > 0.30$).  These results suggest that repositories organized into deeper but less divergent folder structures facilitate correct code implementation, while shallow, highly forked hierarchies impede it.

\subsection{Error Cause Analysis for OpenHands Agents}
In evaluating the OpenHands agents on our benchmark, we identify several recurrent failure modes. Each error category reveals a distinct challenge the agent faced when translating research papers into correct code. We detail these categories below:
% discussing how each issue impeded success and suggesting directions to mitigate them.
\vspace{-1mm}
\paragraph{Unsuccessful Paper Parsing.}
One prominent source of error is the agent’s difficulty in accurately parsing research paper PDFs. Complex layouts and formula-heavy descriptions often lead to incomplete or garbled inputs. For example, mathematical formulas and pseudo-codes frequently fail to extract correctly (missing symbols or mis-ordered elements) which meant the agent lost critical information about the algorithm’s computations. 
\vspace{-2mm}
\paragraph{Incomplete Comprehension of Problem Context.} 
Even when the paper is parsed correctly, the agent often demonstrate a shallow understanding of the implementation instructions. Research papers typically describe algorithms at a high level of abstraction, assuming the reader (or agent) can infer intermediate steps. We observe cases where the model can capture the general idea (e.g., “\textit{apply an attention mechanism}”) but struggle to expand this into concrete code behaviors. Key implementation details that are only implied in the text (such as specific iteration orders, stopping criteria, or parameter initializations) are sometimes omitted in the generated code. This indicates the models are not decomposing the task sufficiently and instead producedan incomplete or overly generic solution.
\vspace{-1mm}

\paragraph{Lack of Robustness in Code Generation.} Many errors originate from intrinsic weaknesses in the LLM’s generated code, including syntax errors, logical mistakes, and incorrect handling of edge cases. These issues reflect limitations in the model’s programming capabilities and its sensitivity to nuanced coding standards. Employing automated code-verification tools and integrating iterative code refinement loops may significantly reduce such errors, improving overall robustness.

% \paragraph{Incomplete Understanding of Dependencies} Failures also frequently occur when the agent does not fully recognize necessary components or dependencies critical for accurate function implementation. This lack of understanding typically results from incomplete or insufficient specification in prompts about required auxiliary functionalities or dependencies. Future benchmarks can address this by providing clearer guidance, including explicit enumeration of required dependencies and components.
\vspace{-1mm}

\paragraph{Cross-file Retrieval Error.} 
The benchmark reveals that agents struggle with retrieving and using information defined across multiple files in a code repository. Often, the function to be implemented relies on constants, helper functions, or class definitions located in other parts of the project. The code agent, with a limited context window, sometimes fail to recall or look up these dependencies. Consequently, it may attempt to redefine a function that already exists elsewhere, use a placeholder value for an unknown constant, or simply omit functionality that depends on unseen parts of the codebase. These mistakes underscore the difficulty of repository-level code generation when not all relevant context fits in the prompt. 

\vspace{-1.5mm}
\paragraph{Policy Errors.}
Another category of failure stems from how the agent handle (or fail to handle) prompt revisions and policy constraints. In some trials, attempts are made to revise the prompt or inject additional context (such as appending external code snippets or altering the task description on the fly). These interventions sometimes trigger LLM’s safety policies or confuse its understanding of the task. In such cases, the model’s performance degrades: it might refuse to continue, produce irrelevant output, or reset its earlier reasoning. 

% \paragraph{Other issues}

% To get a better understanding, we provide a concrete example for each error category, which can be found in Appendix \ref{app:error examples}.
% Additionally, we also provide a statistic analysis for the error distribution of OpenHands on our benchmark. As shown in Figure \ref{fig:error distribution}.

\begin{figure}[t!]
    \centering
    \includegraphics[width=\linewidth]{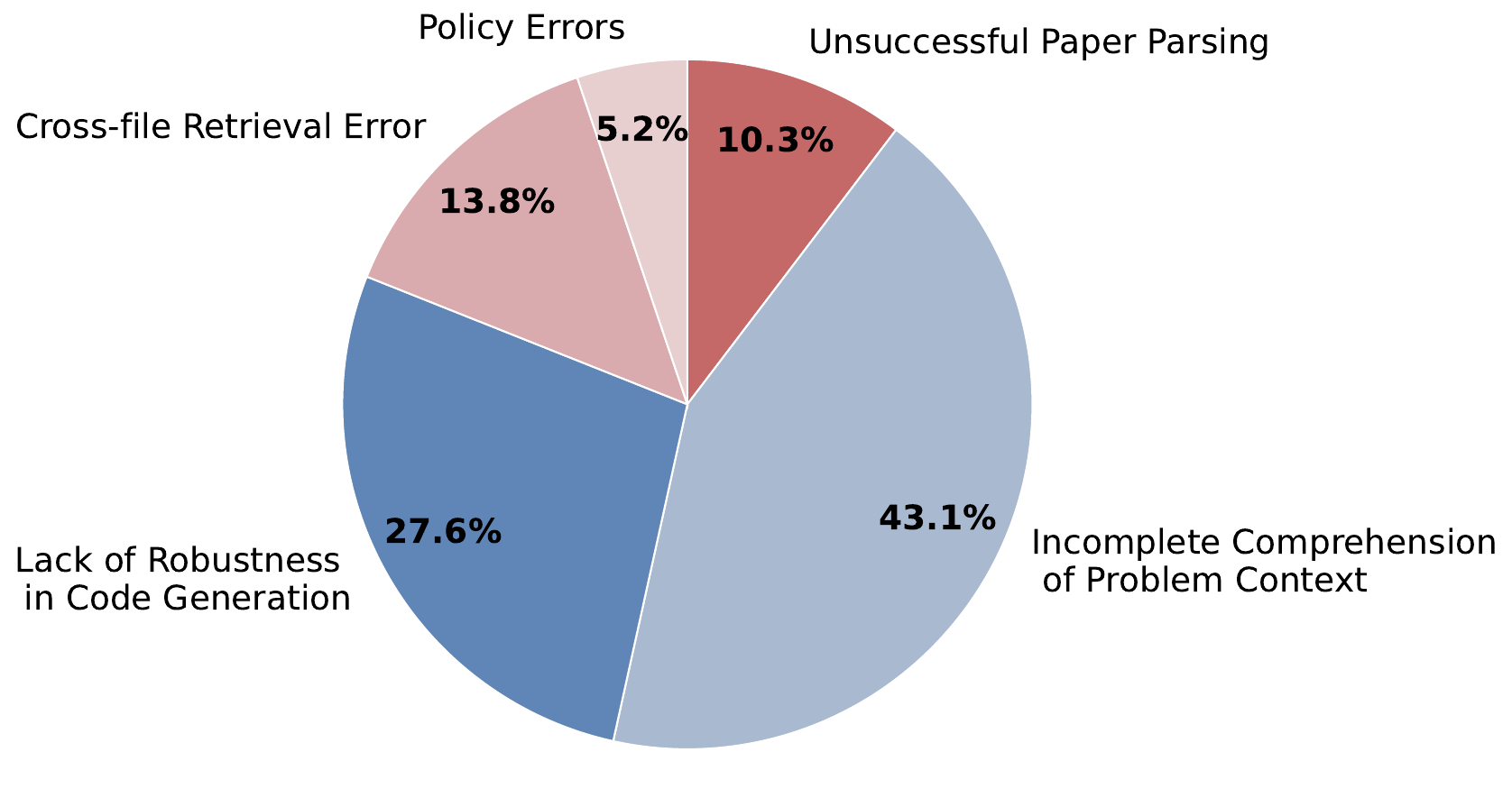}
    \vspace{-6mm}
    \caption{Error distribution of OpenHands on \textsc{\ours}.}
    \label{fig:error distribution}
    \vspace{-5mm}
\end{figure}

To facilitate better understanding, we include a concrete example for each error category in Appendix~\ref{app:error examples}.
Additionally, we provide a statistical analysis of OpenHands' error distribution on our benchmark, as shown in Figure~\ref{fig:error distribution}. The figure reveals a clear hierarchy of failure sources: nearly half (43.1\%) arise from incomplete comprehension of the problem context, while 27.6\% stem from brittle code that breaks under minor input or environment changes. Cross-file retrieval issues account for 13.8\%, underscoring limitations in tracking multi-file dependencies. 
% Improving context understanding and robustness offers the most immediate path for performance gains.

\vspace{-1.5mm}
\subsection{Discussions on Evaluation Methods}
We leverage two automated methods to evaluate the quality of agent-generated code. To further assess both these methods and the code quality itself, we conduct detailed human annotation (Appendix \ref{app:human information}). Table~\ref{tab:model-performance} shows results from unit tests, LLM-as-a-judge, and human judgment. Unit tests achieve the highest accuracy, followed by human evaluation, with LLM-as-a-judge showing the worst performance contrary to our expectations. As LLMs improve, they can generate executable and well-structured code, but there can be multiple correct implementations. The agent may produce correct code that differs from the golden answer, and LLM-as-a-judge, which emphasizes similarity to the reference solution, may fail to recognize its correctness. Human evaluation, while more flexible in accepting varied implementations, tends to be bold. Annotators often label unfamiliar but plausible solutions as ``logically correct'', especially if the implementation does not seem to fully follow the instructions, even when the output is incorrect.

We also observe mismatches between LLM and human judgments. They agreed 62.5\% of the time and directly contradicted each other 9.5\% of the time. Among the four human annotators, two show more frequent disagreement with the LLM (averaging 16\% conflict), mainly on specific papers. In nearly all disagreement cases, humans mark the solution as correct while the LLM do not. The only opposite case involves a code execution failure. LLM-as-a-judge focuses more on surface similarity and details, while humans emphasize overall reasoning, leading to more divergence on complex examples. Still, the agreement exceeds 70\%.

These findings support using both unit tests and LLM-as-a-judge. Unit tests capture \textit{correctness} regardless of implementation strategy, while LLM-as-a-judge measures \textit{alignment} with the reference and instruction \textit{fidelity}. Together, they approximate human judgment well and validate the reliability of our evaluation framework.

% \section{Data Contamination}

% \xd{read MLE-Bench: FAMILIARITY WITH TOP SOLUTIONS}

\vspace{-1.5mm}

\section{Conclusion}
\label{sec:conclusion}
\vspace{-1.5mm}

We present \ours, a benchmark designed to systematically evaluate the LLM agent's ability on reproducing language modeling research.
To ensure an objective evaluation of the code reproduction results, we employ two distinct metrics: the accuracy of unit tests and the distribution of LLM-as-a-judge classifications of generated implementations. 
Experimental results on both standard generation and LLM agent settings reveal the persistent limitations in scientific reasoning and code synthesis of existing models, highlighting critical gaps in agent’s ability to autonomously reproduce scientific research.
In the future, we will focus on automatic or semi-automatic data collection and design more capable agents to improve reproduction outcomes.

% \newpage
\section*{Limitations}

% Since December 2023, a "Limitations" section has been required for all papers submitted to ACL Rolling Review (ARR). This section should be placed at the end of the paper, before the references. The "Limitations" section (along with, optionally, a section for ethical considerations) may be up to one page and will not count toward the final page limit. Note that these files may be used by venues that do not rely on ARR so it is recommended to verify the requirement of a "Limitations" section and other criteria with the venue in question.

To ensure the high quality of our benchmark, the annotation cost is high and scalability is difficult since it requires PhD-level expertise. How to enable automatic or semi–automatic data points collection is an open problem.

\section*{Ethical Considerations}
We developed \ours\ based on research papers from top-tier NLP conference proceedings and their publicly available code repositories. This project has been classified as exempt by our Institutional Review Board (IRB). All human annotations and evaluations were conducted by our co-authors who are researchers with substantial NLP research experience. The systems trained on our dataset are intended to augment—not replace—human decision-making in scientific research.

% \section*{Acknowledgments}

% This document has been adapted
% by Steven Bethard, Ryan Cotterell and Rui Yan
% from the instructions for earlier ACL and NAACL proceedings, including those for
% ACL 2019 by Douwe Kiela and Ivan Vuli\'{c},
% NAACL 2019 by Stephanie Lukin and Alla Roskovskaya,
% ACL 2018 by Shay Cohen, Kevin Gimpel, and Wei Lu,
% NAACL 2018 by Margaret Mitchell and Stephanie Lukin,
% Bib\TeX{} suggestions for (NA)ACL 2017/2018 from Jason Eisner,
% ACL 2017 by Dan Gildea and Min-Yen Kan,
% NAACL 2017 by Margaret Mitchell,
% ACL 2012 by Maggie Li and Michael White,
% ACL 2010 by Jing-Shin Chang and Philipp Koehn,
% ACL 2008 by Johanna D. Moore, Simone Teufel, James Allan, and Sadaoki Furui,
% ACL 2005 by Hwee Tou Ng and Kemal Oflazer,
% ACL 2002 by Eugene Charniak and Dekang Lin,
% and earlier ACL and EACL formats written by several people, including
% John Chen, Henry S. Thompson and Donald Walker.
% Additional elements were taken from the formatting instructions of the \emph{International Joint Conference on Artificial Intelligence} and the \emph{Conference on Computer Vision and Pattern Recognition}.

\bibliography{custom}

\appendix

\appendix

\onecolumn

% \section{Technical Appendices and Supplementary Material}
% Technical appendices with additional results, figures, graphs and proofs may be submitted with the paper submission before the full submission deadline (see above), or as a separate PDF in the ZIP file below before the supplementary material deadline. There is no page limit for the technical appendices.

\section{Prompt for LLM-as-a-Judge evaluation}
\label{appendix:extract_idea}

\begin{tcolorbox}[title=Prompt for LLM-as-a-Judge Evaluation, left=2mm,right=1mm,top=0mm, bottom=0mm,colback=white]
\begin{lstlisting}[style=plain]
Instruction: {instruction}

You are an expert NLP software engineer tasked with evaluating the correctness of a function implementation by comparing two code artifacts:

- Golden Reference ({golden file}):
  {golden content}

- Agent Implementation ({goal file}):
  {goal content}

Instructions:
1. Examine both implementations in detail, focusing on:
   - Logical correctness relative to the specification provided above
   - Handling of edge cases and error conditions
   - Subtle deviations such as off-by-one errors or missing checks

2. Classify your judgment into exactly one of the following categories:
   1. Incorrect Logic: the core algorithm deviates from the specification and produces wrong results
   2. Logic Correct but Subtle Errors: the main algorithm matches the specification, but there are other implementation mistakes or omissions
   3. Completely Correct: the implementation is fully faithful to the specification with no errors

3. For the chosen category, provide a concise rationale with two to four bullet points illustrating the key discrepancies or confirmations

Output Format (JSON):
{
  category: <1 | 2 | 3>,
  rationale: [
    First key point ...,
    Second key point ...
  ]
}
\end{lstlisting}
\end{tcolorbox}

% \begin{tcolorbox}[appendixbox]

% Instruction: \{instruction\}

% You are an expert NLP software engineer tasked with evaluating the correctness of a function implementation by comparing two code artifacts:

% - Golden Reference (\{golden file\}):
%   \{golden content\}

% - Agent Implementation (\{goal file\}):
%   \{goal content\}

% Instructions:
% 1. Examine both implementations in detail, focusing on:
%    - Logical correctness relative to the specification provided above
%    - Handling of edge cases and error conditions
%    - Subtle deviations such as off-by-one errors or missing checks

% 2. Classify your judgment into exactly one of the following categories:
%    1. Incorrect Logic: the core algorithm deviates from the specification and produces wrong results
%    2. Logic Correct but Subtle Errors: the main algorithm matches the specification, but there are other implementation mistakes or omissions
%    3. Completely Correct: the implementation is fully faithful to the specification with no errors

% 3. For the chosen category, provide a concise rationale with two to four bullet points illustrating the key discrepancies or confirmations

% Output Format (JSON):
% {
%   category: <1 | 2 | 3>,
%   rationale: [
%     First key point…,
%     Second key point…
%   ]
% }

% \end{tcolorbox}

\section{Prompt for Standard Prompting}
\label{appendix:prompt-nov}

\begin{tcolorbox}[title=Prompt for Standard Prompting, left=2mm,right=1mm,top=0mm, bottom=0mm,colback=white]
\begin{lstlisting}[style=plain]
You are a code assistant.
Below is the entire Python source file.
Please implement only the function/method named <method_name>.
Return only its def line and indented body--no fences or explanations.

=== FILE BEGIN ===
<full_source_code>
=== FILE END ===

Paper (JSON):
<paper_json>

Instruction:
<instruction>

Related code for reference:       
# Path: <relative_path_1>
<retrieval_content_1>
# Path: <relative_path_2>
<retrieval_content_2>
\end{lstlisting}
\end{tcolorbox}

\newpage
% \begin{tcolorbox}[appendixbox]
% \noindent
% You are a code assistant.
% Below is the entire Python source file.
% Please implement only the function/method named <method_name>.
% Return only its def line and indented body—no fences or explanations.

% === FILE BEGIN ===
% <full_source_code>
% === FILE END ===

% Paper (JSON):
% <paper_json>

% Instruction:
% <instruction>

% Related code for reference:       
% # Path: <relative_path_1>
% <retrieval_content_1>
% # Path: <relative_path_2>
% <retrieval_content_2>
% …

% \end{tcolorbox}

\section{Paper Categories in \ours}
\begin{table*}[h]
    \centering
    \small
    \begin{tabular}{p{0.21\linewidth}|p{0.73\linewidth}}
        \toprule
        \textbf{Category} & \textbf{Definition} \\
        \midrule
        \textit{Feature Learning \& \newline Representation} & \multirow{2}{*}{Creating and refining vector representation of texts.} \\
        \midrule
        \textit{Neural Network \newline Architectures} & \multirow{2}{*}{Designing building blocks within neural networks.} \\
        \midrule
        \textit{Prompt Engineering \& \newline Instruction Tuning} & \multirow{2}{*}{Crafting prompts or fine‑tuning models to control and optimize model behaviors.}\\
        \midrule
        \textit{Information Extraction} & Extracting structured knowledge from unstructured texts. \\
        \midrule
        \textit{Data Augmentation} & Augmenting training samples with curated strategies.  \\
        \midrule
        \textit{Training Objectives \& \newline Optimization} & \multirow{2}{*}{Designing loss functions or optimization algorithms to govern model training.} \\
        \midrule
        \textit{Decoding \& \newline Search Strategies} & \multirow{2}{*}{Employing inference-time algorithms for decoding and search in text generation.} \\
        \midrule
        \textit{Evaluation Metrics} & Calculating quantitative measures of text generation results. \\
        \midrule
        \textit{Interpretability \& \newline Explainability} & \multirow{2}{*}{Focusing on techniques that illuminate model internals and decision rationales.} \\
        \bottomrule
    \end{tabular}
        \caption{Paper categories within the \textsc{\ours} benchmark.}
    \label{tab:categories}
\end{table*}

\section{Human Annotator Profile}
\label{app:human information}
To conduct the manual annotation evaluation, we first draw a random sample of 40 papers.  We then recruit a panel of 30 subject-matter specialists drawn from a variety of institutions and including several faculty members to annotate their papers.  Each annotated paper is examined independently by three different experts, ensuring the correctness of unit tests. On average, completing one paper annotation requires roughly five hours. 

\paragraph{Qualification Standards.}  
All annotators are selected to satisfy the reviewer expectations of premier NLP and machine learning venues. Each expert met at least \emph{two} of the following conditions:

\begin{itemize}
    \item holds a Ph.D.\ or has authored multiple peer-reviewed publications in a relevant discipline;
    \item has published a minimum of two \emph{first-authored} papers in top-tier conferences or journals (AAAI, NeurIPS, ICML, ICLR, ACL, EMNLP, NAACL, etc.) within the past five years;
    \item has served as a reviewer for one of these venues, or has comparable research standing as demonstrated by citation metrics and scholarly record.
\end{itemize}

\paragraph{Evaluation Protocol.}  
During the manual evaluation stage, each annotator are provided with the uniform evaluate criteria, and are be required to write a brief justification of their scoring. On average, it takes around ten minutes to complete a code evaluation.

\newpage

\section{Example of \ours}
\label{app:dataset_example}

\begin{tcolorbox}[title=Example of the DPO Paper in \ours, left=2mm,right=1mm,top=0mm, bottom=0mm,colback=white]
\begin{lstlisting}[style=plain]
Structure of data folder:
- 1-DPO
    - direct-preference-optimization (main code repository)
    - Golden_files (reference implementation files)
    - Dockerfile (defines Docker environment for unit-test evaluation)
    - info.json (metadata and implementation details)

Content of info.json:
{
    "instance_id": 1,
    "paper_name": "Direct Preference Optimization: Your Language Model is Secretly a Reward Model",
    "folder_name": "1-DPO",
    "paper_url": "https://arxiv.org/pdf/2305.18290",
    "year": 2023,
    "repo_url": "https://github.com/eric-mitchell/direct-preference-optimization",
    "repo_folder_name": "direct-preference-optimization",
    "implementations": [
        {
            "instruction": "Implement the preference_loss function in trainers.py based on the DPO loss mentioned in the paper and the code repository. You may ignore the following parameters: ipo, reference_free and label_smoothing.",
            "index": 1,
            "category": "Training Objectives & Optimization Techniques",
            "goal_file": "trainers.py",
            "goal_function": "preference_loss",
            "class_name": "",
            "golden_file": "golden_files/trainers_golden.py",
            "retrieval_context": [],
            "unit_test_file": "unit_test/unit_test_1.py"
        }
    ]
}


\end{lstlisting}
\end{tcolorbox}

\newpage

\section{Example of Incorrect Implementations by OpenHands}
\label{app:error examples}
\paragraph{Unsuccessful Paper Parsing.}
Blow is the agent implementation of Native Sparse Attention~\citep{DBLP:journals/corr/abs-2502-11089}. A comparison with the original algorithmic design reveals that the agent’s implementation fails to capture the core algorithmic flow described in the paper, indicating a fundamental failure in parsing and understanding the intended methodology.

\newpage

\begin{tcolorbox}[title=OpenHands' Implementation, left=2mm,right=1mm,top=0mm, bottom=0mm,colback=white]
\begin{lstlisting}[style=plain]
def nsa(q: torch.Tensor,
    k: torch.Tensor,
    v: torch.Tensor,
    g_slc: torch.Tensor,
    g_swa: torch.Tensor,
    block_indices: torch.LongTensor,
    block_counts: Optional[Union[torch.LongTensor, int]] = None,
    block_size: int = 64,
    window_size: int = 0,
    scale: Optional[float] = None,
    cu_seqlens: Optional[torch.LongTensor] = None,
    head_first: bool = False) -> torch.Tensor:
    r"""
    Args:
        q (torch.Tensor): Queries of shape `[B, T, HQ, K]` if `head_first=False` else `[B, HQ, T, K]`.
        k (torch.Tensor): Keys of shape `[B, T, H, K]` if `head_first=False` else `[B, H, T, K]`.
            GQA is enforced here. The ratio of query heads (HQ) to key/value heads (H) must be a power of 2 and >=16.
        v (torch.Tensor): Values of shape `[B, T, H, V]` if `head_first=False` else `[B, H, T, V]`.
        g_slc (torch.Tensor): Gate score for selected attention of shape `[B, T, HQ]` if  `head_first=False` else `[B, HQ, T]`.
        g_swa (torch.Tensor): Gate score for sliding attentionof shape `[B, T, HQ]` if  `head_first=False` else `[B, HQ, T]`.
        block_indices (torch.LongTensor): Block indices of shape `[B, T, H, S]` if `head_first=False` else `[B, H, T, S]`.
            `S` is the maximum number of selected blocks for each query token, which is set to 16 in the paper.
        block_counts (Union[torch.LongTensor, int]): Number of selected blocks for each token.  If a tensor is provided, with shape `[B, T, H]` if `head_first=True` else `[B, T, H]`, each token can select the same number of blocks.If not provided, it will default to `S`, Default: `None`.
        block_size (int): Selected block size. Default: 64.
        window_size (int): Sliding window size. Default: 0.
        scale (Optional[int]): Scale factor for attention scores.
            If not provided, it will default to `1 / sqrt(K)`. Default: `None`.
        cu_seqlens (torch.LongTensor): Cumulative sequence lengths of shape `[N+1]` used for variable-length training,
            consistent with the FlashAttention API.
        head_first (Optional[bool]):
            Whether the inputs are in the head-first format. Default: `False`.
    Returns:
        o (torch.Tensor):
            Outputs of shape `[B, T, HQ, V]` if `head_first=False` else `[B, HQ, T, V]`.
    """

    # Determine scale if not provided
    if scale is None:
        scale = 1 / math.sqrt(q.size(-1))

    # Compute scaled dot-product attention
    q = q * scale

    # Hierarchical Sparse Strategy: Coarse-grained token compression
    # Assuming g_slc and g_swa are gate scores for selected and sliding attention
    # Apply gate scores to queries
    q_slc = q * g_slc.unsqueeze(-1)
    q_swa = q * g_swa.unsqueeze(-1)

    # Fine-grained token selection
    # Compute attention scores for selected blocks
    attn_scores_slc = torch.einsum('bthd,bshd->bths', q_slc, k)
    attn_scores_swa = torch.einsum('bthd,bshd->bths', q_swa, k)

    # Apply softmax to get attention probabilities
    attn_probs_slc = F.softmax(attn_scores_slc, dim=-1)
    attn_probs_swa = F.softmax(attn_scores_swa, dim=-1)

    # Compute attention outputs
    o_slc = torch.einsum('bths,bshd->bthd', attn_probs_slc, v)
    o_swa = torch.einsum('bths,bshd->bthd', attn_probs_swa, v)

    # Combine outputs from selected and sliding attention
    o = o_slc + o_swa

    return o
\end{lstlisting}
\end{tcolorbox}

\paragraph{Incomplete Comprehension of Problem Context.} 

OpenHands' implementation compared with reference implementation and the backbone of OpenHands is o4-mini. Parameter $\texttt{beta}$ is not used correctly in the OpenHands's implementation.

\begin{tcolorbox}[title=OpenHands' Implementation, left=2mm,right=1mm,top=0mm, bottom=0mm,colback=white]
\begin{lstlisting}[style=plain]
def preference_loss(policy_chosen_logps: torch.FloatTensor,
                    policy_rejected_logps: torch.FloatTensor,
                    reference_chosen_logps: torch.FloatTensor,
                    reference_rejected_logps: torch.FloatTensor,
                    beta: float,
                    label_smoothing: float = 0.0,
                    ipo: bool = False,
                    reference_free: bool = False) -> Tuple[torch.FloatTensor, torch.FloatTensor, torch.FloatTensor]:
    """Compute the DPO loss for a batch of policy and reference model log probabilities.

    Args:
        policy_chosen_logps: Log probabilities of the policy model for the chosen responses. Shape: (batch_size,)
        policy_rejected_logps: Log probabilities of the policy model for the rejected responses. Shape: (batch_size,)
        reference_chosen_logps: Log probabilities of the reference model for the chosen responses. Shape: (batch_size,)
        reference_rejected_logps: Log probabilities of the reference model for the rejected responses. Shape: (batch_size,)
        beta: Temperature parameter for the DPO loss, typically something in the range of 0.1 to 0.5. We ignore the reference model as beta -> 0.
        label_smoothing: conservativeness for DPO loss, which assumes that preferences are noisy (flipped with probability label_smoothing)
        ipo: If True, use the IPO loss instead of the DPO loss.
        reference_free: If True, we ignore the _provided_ reference model and implicitly use a reference model that assigns equal probability to all responses.

    Returns:
        A tuple of three tensors: (losses, chosen_rewards, rejected_rewards).
        The losses tensor contains the DPO loss for each example in the batch.
        The chosen_rewards and rejected_rewards tensors contain the rewards for the chosen and rejected responses, respectively.
    """
if reference_free:
        reference_chosen_logps = torch.zeros_like(reference_chosen_logps)
        reference_rejected_logps = torch.zeros_like(reference_rejected_logps)
    chosen_rewards = (policy_chosen_logps - reference_chosen_logps) / beta
    rejected_rewards = (policy_rejected_logps - reference_rejected_logps) / beta
    differences = chosen_rewards - rejected_rewards
    losses = -F.logsigmoid(differences)
    return losses, chosen_rewards, rejected_rewards


\end{lstlisting}
\end{tcolorbox}

\newpage

\begin{tcolorbox}[title=Reference Implementation, left=2mm,right=1mm,top=0mm, bottom=0mm,colback=white]
\begin{lstlisting}[style=plain]
def preference_loss(policy_chosen_logps: torch.FloatTensor,
                    policy_rejected_logps: torch.FloatTensor,
                    reference_chosen_logps: torch.FloatTensor,
                    reference_rejected_logps: torch.FloatTensor,
                    beta: float,
                    label_smoothing: float = 0.0,
                    ipo: bool = False,
                    reference_free: bool = False) -> Tuple[torch.FloatTensor, torch.FloatTensor, torch.FloatTensor]:
    """Compute the DPO loss for a batch of policy and reference model log probabilities.

    Args:
        policy_chosen_logps: Log probabilities of the policy model for the chosen responses. Shape: (batch_size,)
        policy_rejected_logps: Log probabilities of the policy model for the rejected responses. Shape: (batch_size,)
        reference_chosen_logps: Log probabilities of the reference model for the chosen responses. Shape: (batch_size,)
        reference_rejected_logps: Log probabilities of the reference model for the rejected responses. Shape: (batch_size,)
        beta: Temperature parameter for the DPO loss, typically something in the range of 0.1 to 0.5. We ignore the reference model as beta -> 0.
        label_smoothing: conservativeness for DPO loss, which assumes that preferences are noisy (flipped with probability label_smoothing)
        ipo: If True, use the IPO loss instead of the DPO loss.
        reference_free: If True, we ignore the _provided_ reference model and implicitly use a reference model that assigns equal probability to all responses.

    Returns:
        A tuple of three tensors: (losses, chosen_rewards, rejected_rewards).
        The losses tensor contains the DPO loss for each example in the batch.
        The chosen_rewards and rejected_rewards tensors contain the rewards for the chosen and rejected responses, respectively.
    """
    pi_logratios = policy_chosen_logps - policy_rejected_logps
    ref_logratios = reference_chosen_logps - reference_rejected_logps

    if reference_free:
        ref_logratios = 0

    logits = pi_logratios - ref_logratios  # also known as h_{\pi_\theta}^{y_w,y_l}

    if ipo:
        losses = (logits - 1/(2 * beta)) ** 2  # Eq. 17 of https://arxiv.org/pdf/2310.12036v2.pdf
    else:
        # Eq. 3 https://ericmitchell.ai/cdpo.pdf; label_smoothing=0 gives original DPO (Eq. 7 of https://arxiv.org/pdf/2305.18290.pdf)
        losses = -F.logsigmoid(beta * logits) * (1 - label_smoothing) - F.logsigmoid(-beta * logits) * label_smoothing

    chosen_rewards = beta * (policy_chosen_logps - reference_chosen_logps).detach()
    rejected_rewards = beta * (policy_rejected_logps - reference_rejected_logps).detach()

    return losses, chosen_rewards, rejected_rewards

\end{lstlisting}
\end{tcolorbox}

\newpage

\paragraph{Lack of Robustness in Code Generation.} Below is an example of OpenHands' implementation compared with reference implementation and the backbone of OpenHands is GPT-4o. There is two case in the reference implementation determined by parameter $\texttt{if\_tdpo2}$ whether to use method TDPO2. However, the implementation by OpenHands neglects to consider these cases.

\begin{tcolorbox}[title=OpenHands' Implementation, left=2mm,right=1mm,top=0mm, bottom=0mm,colback=white]
\begin{lstlisting}[style=plain]
def tdpo_loss(chosen_logps_margin: torch.FloatTensor,
              rejected_logps_margin: torch.FloatTensor,
              chosen_position_kl: torch.FloatTensor,
              rejected_position_kl: torch.FloatTensor,
              beta: float, alpha: float = 0.5, if_tdpo2: bool = True) -> Tuple[torch.FloatTensor, torch.FloatTensor, torch.FloatTensor]:
    """Compute the TDPO loss for a batch of policy and reference model log probabilities.

    Args:
        chosen_logps_margin: The difference of log probabilities between the policy model and the reference model for the chosen responses. Shape: (batch_size,)
        rejected_logps_margin: The difference of log probabilities between the policy model and the reference model for the rejected responses. Shape: (batch_size,)
        chosen_position_kl: The difference of sequential kl divergence between the policy model and the reference model for the chosen responses. Shape: (batch_size,)
        rejected_position_kl: The difference of sequential kl divergence between the policy model and the reference model for the rejected responses. Shape: (batch_size,)
        beta: Temperature parameter for the TDPO loss, typically something in the range of 0.1 to 0.5. We ignore the reference model as beta -> 0.
        alpha: Temperature parameter for the TDPO loss, used to adjust the impact of sequential kl divergence.
        if_tdpo2: Determine whether to use method TDPO2, default is True; if False, then use method TDPO1.

    Returns:
        A tuple of two tensors: (losses, rewards).
        The losses tensor contains the TDPO loss for each example in the batch.
        The rewards tensors contain the rewards for response pair.
    """
    # Calculate the rewards using the Bradley-Terry model
    chosen_rewards = torch.sigmoid(chosen_logps_margin / beta - alpha * chosen_position_kl)
    rejected_rewards = torch.sigmoid(rejected_logps_margin / beta - alpha * rejected_position_kl)

    # Calculate the losses as the negative log likelihood of the rewards
    losses = -torch.log(chosen_rewards) - torch.log(1 - rejected_rewards)

    return losses, chosen_rewards, rejected_rewards

\end{lstlisting}
\end{tcolorbox}

\newpage

\begin{tcolorbox}[title=Reference Implementation, left=2mm,right=1mm,top=0mm, bottom=0mm,colback=white]
\begin{lstlisting}[style=plain]
def tdpo_loss(chosen_logps_margin: torch.FloatTensor,
              rejected_logps_margin: torch.FloatTensor,
              chosen_position_kl: torch.FloatTensor,
              rejected_position_kl: torch.FloatTensor,
              beta: float, alpha: float = 0.5, if_tdpo2: bool = True) -> Tuple[torch.FloatTensor, torch.FloatTensor, torch.FloatTensor]:
    """Compute the TDPO loss for a batch of policy and reference model log probabilities.

    Args:
        chosen_logps_margin: The difference of log probabilities between the policy model and the reference model for the chosen responses. Shape: (batch_size,)
        rejected_logps_margin: The difference of log probabilities between the policy model and the reference model for the rejected responses. Shape: (batch_size,)
        chosen_position_kl: The difference of sequential kl divergence between the policy model and the reference model for the chosen responses. Shape: (batch_size,)
        rejected_position_kl: The difference of sequential kl divergence between the policy model and the reference model for the rejected responses. Shape: (batch_size,)
        beta: Temperature parameter for the TDPO loss, typically something in the range of 0.1 to 0.5. We ignore the reference model as beta -> 0.
        alpha: Temperature parameter for the TDPO loss, used to adjust the impact of sequential kl divergence.
        if_tdpo2: Determine whether to use method TDPO2, default is True; if False, then use method TDPO1.

    Returns:
        A tuple of two tensors: (losses, rewards).
        The losses tensor contains the TDPO loss for each example in the batch.
        The rewards tensors contain the rewards for response pair.
    """

    chosen_values = chosen_logps_margin + chosen_position_kl
    rejected_values = rejected_logps_margin + rejected_position_kl

    chosen_rejected_logps_margin = chosen_logps_margin - rejected_logps_margin


    if not if_tdpo2:
        logits = chosen_rejected_logps_margin - (rejected_position_kl - chosen_position_kl)    # tdpo1
    else:
        logits = chosen_rejected_logps_margin - alpha * (rejected_position_kl - chosen_position_kl.detach())  # tdpo2
    losses = -F.logsigmoid(beta * logits)

    chosen_rewards = beta * chosen_values.detach()
    rejected_rewards = beta * rejected_values.detach()

    return losses, chosen_rewards, rejected_rewards

\end{lstlisting}
\end{tcolorbox}

\newpage

\paragraph{Cross-file Retrieval Error.}

% In the reference implementation of following example, the \texttt{sumpo_loss} function is defined as a class method and therefore depends on instance attributes such as \texttt{self.gamma_beta_ratio} and \texttt{self.beta}. In the agent implementation, however, the function is declared outside the class scope, making these attributes inaccessible. As a result, \texttt{sumpo_loss} is executed without the required parameters, yielding an ill-defined—or “ungrounded”—implementation.

In the following example, \texttt{simpo\_loss} depends on class-level attributes (e.g., \texttt{self.gamma\_beta\_ratio} and \texttt{self.beta}) defined outside the function body. As a result, the agent implementation fails to access these key parameters from \texttt{self}, leading to an incorrect implementation of the \texttt{simpo\_loss} function.

\begin{tcolorbox}[title=OpenHands' Implementation, left=2mm,right=1mm,top=0mm, bottom=0mm,colback=white]
\begin{lstlisting}[style=plain]
    def simpo_loss(
        self,
        policy_chosen_logps: torch.FloatTensor,
        policy_rejected_logps: torch.FloatTensor,
    ) -> Tuple[torch.FloatTensor, torch.FloatTensor, torch.FloatTensor]:
        """Compute the SimPO loss for a batch of policy model log probabilities.

        Args:
            policy_chosen_logps: Log probabilities of the policy model for the chosen responses. Shape: (batch_size,)
            policy_rejected_logps: Log probabilities of the policy model for the rejected responses. Shape: (batch_size,)

        Returns:
            A tuple of three tensors: (losses, chosen_rewards, rejected_rewards).
            The losses tensor contains the SimPO loss for each example in the batch.
            The chosen_rewards and rejected_rewards tensors contain the rewards for the chosen and rejected responses, respectively.
        """

        # Calculate rewards using average log probabilities
        chosen_rewards = policy_chosen_logps.mean(dim=-1)
        rejected_rewards = policy_rejected_logps.mean(dim=-1)

        # Implement the Bradley-Terry objective with a target reward margin
        target_margin = 1.0  # This can be a hyperparameter
        margin = chosen_rewards - rejected_rewards - target_margin
        losses = -F.logsigmoid(margin)

        return losses, chosen_rewards, rejected_rewards
\end{lstlisting}
\end{tcolorbox}

\newpage

\begin{tcolorbox}[title=Reference Implementation, left=2mm,right=1mm,top=0mm, bottom=0mm,colback=white]
\begin{lstlisting}[style=plain]
    def simpo_loss(
        self,
        policy_chosen_logps: torch.FloatTensor,
        policy_rejected_logps: torch.FloatTensor,
    ) -> Tuple[torch.FloatTensor, torch.FloatTensor, torch.FloatTensor]:
        """Compute the SimPO loss for a batch of policy model log probabilities.

        Args:
            policy_chosen_logps: Log probabilities of the policy model for the chosen responses. Shape: (batch_size,)
            policy_rejected_logps: Log probabilities of the policy model for the rejected responses. Shape: (batch_size,)

        Returns:
            A tuple of three tensors: (losses, chosen_rewards, rejected_rewards).
            The losses tensor contains the SimPO loss for each example in the batch.
            The chosen_rewards and rejected_rewards tensors contain the rewards for the chosen and rejected responses, respectively.
        """
        pi_logratios = policy_chosen_logps - policy_rejected_logps
        logits = pi_logratios - self.gamma_beta_ratio

        if self.loss_type == "sigmoid":
            losses = (
                -F.logsigmoid(self.beta * logits) * (1 - self.label_smoothing)
                - F.logsigmoid(-self.beta * logits) * self.label_smoothing
            )
        elif self.loss_type == "hinge":
            losses = torch.relu(1 - self.beta * logits)
        else:
            raise ValueError(
                f"Unknown loss type: {self.loss_type}. Should be one of ['sigmoid', 'hinge']"
            )

        chosen_rewards = self.beta * policy_chosen_logps
        rejected_rewards = self.beta * policy_rejected_logps

        return losses, chosen_rewards, rejected_rewards

\end{lstlisting}
\end{tcolorbox}

\newpage

\paragraph{Policy Errors.} Below is an OpenHands implementation using GPT-4o as its backbone. Notably, the target function remains unimplemented:

\begin{tcolorbox}[title=Agent implementation, left=2mm,right=1mm,top=0mm, bottom=0mm,colback=white]
\begin{lstlisting}[style=plain]
    def info_nce_loss(self, features):
        """
        Compute the InfoNCE loss for a batch of features.

        Args:
            features (torch.Tensor): Normalized feature representations from the encoder.
                Shape: (batch_size * n_views, feature_dim).
                It is assumed that features from different augmented views of the same image
                are stacked along the batch dimension.

        Returns:
            A tuple containing:
                - logits (torch.Tensor): Similarity scores for positive and negative pairs.
                  Shape: (batch_size * n_views, 1 + num_negatives).
                  Each row corresponds to one positive pair and multiple negative pairs.
                - labels (torch.Tensor): Ground truth labels where the first entry is the positive.
                  Shape: (batch_size * n_views,). All entries are 0 since positive is first.
        """


        return logits, labels

\end{lstlisting}
\end{tcolorbox}

The corresponding OpenHands logs below indicate that this run resulted in a \texttt{BadRequestError}, likely because prompt revisions made by OpenHands triggered the underlying LLM’s safety mechanisms.

\begin{tcolorbox}[title=Log Messages, left=2mm,right=1mm,top=0mm, bottom=0mm,colback=white]
\begin{lstlisting}[style=plain]
{"id": 84, "timestamp": "2025-05-16T18:44:29.250531", "source": "environment", "message": "", "observation": "agent_state_changed", "content": "", "extras": {"agent_state": "error", "reason": "BadRequestError: litellm.BadRequestError: OpenAIException - Invalid prompt: your prompt was flagged as potentially violating our usage policy. Please try again with a different prompt: https://platform.openai.com/docs/guides/reasoning#advice-on-prompting"}}
\end{lstlisting}
\end{tcolorbox}

\section{Logistic Regression Analysis}
\label{app:logistic_analysis}
Logistic regression analysis of repository structure and model type on implementation success is shown in Table \ref{tab:logit-results}.
\begin{table*}[h]
\centering
% \small
\begin{tabular}{lrrrrrr}
\toprule
Variable & coef  &  std err & z &  P$>$|z| & [0.025 & 0.975]  \\
\midrule
\texttt{const}  &  -0.4313 &    0.330 &  -1.306  &  0.192     & -1.079   &  0.216  \\
\texttt{avg\_dir\_depth}  &  0.8049 &    0.249 &   3.232  &  0.001     &  0.317   &  1.293  \\
\texttt{avg\_branch\_factor}  &  -0.6750 &    0.225 &  -2.995  &  0.003     & -1.117   & -0.233  \\
\texttt{directory\_imbalance} &   0.2426 &    0.226 &   1.071  &  0.284     & -0.201   &  0.686  \\
\texttt{goal\_function\_length}    &  -0.3478 &    0.225 &  -1.547  &  0.122     & -0.788   &  0.093  \\
\texttt{model\_name\_gpt4o}   &  -0.4458 &    0.475 &  -0.939  &  0.348     & -1.376   &  0.484  \\
\texttt{model\_name\_o4mini}  &   0.2807 &    0.476 &   0.590  &  0.555     & -0.653   &  1.214  \\
\bottomrule
\end{tabular}
\caption{Logistic regression analysis of repository structure and model type on implementation success.}
% \vspace{-4mm}
\label{tab:logit-results}
\end{table*}

\end{document}